\documentclass[twocol]{ametsocV6.1}

\usepackage{float}

\title{SFY --- A lightweight, high-frequency and phase-resolving wave-buoy for coastal waters}

\authors{
  Gaute Hope,\aff{a}\correspondingauthor{Gaute Hope, gauteh@met.no / gaute.hope@gmail.com}
  Torunn Irene Seldal,\aff{a,b}
  Jean Rabault,\aff{a}
  Helge Thomas Bryhni,\aff{b}
  Patrik Bohlinger,\aff{a}
  Jan-Victor Björkqvist,\aff{a,e}
  Tor Nordam,\aff{c,d}
  Atle Kleven,\aff{d}
  Arsalan Mostaani,\aff{c}
  Birgitte Rugaard Furevik,\aff{a,b}
  Lars Robert Hole,\aff{a}
  Roger Storvik,\aff{a}
  Øyvind Breivik\aff{a,b}
}

\affiliation{
  \aff{a}{Norwegian Meteorological Institute}\\
  \aff{b}{University of Bergen}\\
  \aff{c}{Department of Physics, NTNU}\\
  \aff{d}{SINTEF Ocean}\\
  \aff{e}{Finnish Meteorological Institute}
}

\abstract{Small lightweight wave buoys, SFYs, designed to operate near the coast, have been developed. The buoys are designed to record and transmit the full time series of surface acceleration at $52$ Hz. The buoy uses the cellular network to transfer data and position (up to $80$ km from the base station). This reduces costs and increases band-width. The low cost and low weight permits the buoys to be deployed easily, and in arrays in areas where satellite and wave models struggle to resolve wave and current interaction. The buoys are tested in a wave-flume, the open water and in the breaking waves of the surf. The conditions range from calm to significant wave heights exceeding $7$ m and crashing breakers with accelerations exceeding $10g$. The high sample rate captures the impulse of breaking waves, and allows them to be studied in detail. Breaking waves are measured and quantified in the open water. We measure breaking waves in the surf and recover the trajectory of waves breaking in the field to a higher degree than previously done. The time series of surface elevation, and accurate positioning, permits the signal of adjacent buoys to be correlated in a coherent phase-resolved way. Finally, we offer an explanation and solution for the ubiquitous low-frequency noise in IMU-based buoys and discuss necessary sampling and design to measure in areas of breaking waves.}

\begin{document}

\maketitle

%
%
%
%
%
%

%

\section{Introduction}
\label{sec:introduction}

Wave breaking is the primary process by which  energy, momentum and gases are exchanged between the air and the sea. It is the limiting factor in wave growth and helps transfer energy through the wave spectrum from shorter to longer waves \citep{melville1996RoleSurfaceWaveBreaking,babanin2009BreakingOceanSurface,deike2022MassTransferOcean}. The forces induced by breaking waves are one of the dimensioning factors in marine and nautical engineering, and their effect on coastlines is profound. Most maritime activity happens close to the coast, with more than 60 \% of humans living near the coast \citep{boehm2017OceansPerilGrand}. Estimates further suggest that $80$ \% of all search and rescue cases are found within $20$ km of the shoreline \citep{breivik2008OperationalSearchRescue}.
Winds, waves, and currents are heavily modified and influenced by the coastline where waves and currents interact strongly \citep{halsne2022ResolvingRegionsKnown,halsne2023WaveModulationStrong,halsne2024WaveModulationStrong}. Consequently, the regions with the most complex wave conditions typically hold the greatest societal importance and are also the most difficult to sample.

Of the many ways to measure waves, in situ heave buoys and wave staffs are the most popular \citep{holthuijsen2010WavesOceanicCoastal}. There exists a range of high-quality commercial moored and drifting wave buoys [see, e.g., the buoys by Datawell \citep{datawellbv2023DatawellWaveriderManual}, SOFAR Ocean \citep{raghukumar2019PerformanceCharacteristicsSpotter}, Aanderaa \citep{saetre2023DirectionalWaveMeasurements}, and LainePoiss \citep{alari2022LainePoissLightweightIceResistant}]. These buoys are designed for operational application and therefore have a need to function robustly for long stretches of times. They usually report statistical properties, like significant wave height, and in some cases time series up to a few Hz. Often the time series are only available onboard, and not over telemetry. However, the last decade has seen a small revolution in low-cost, open-source wave-measuring devices  \citep{thomson2023DevelopmentTestingMicroSWIFTa,rabault2020OpenSourceVersatile,feddersen2023WavedrifterLowcostIMUbased,collins2024PerformanceMooredGPS}. These designs generally do not seek to have the same robustness as commercial buoys, but can measure in quite different conditions. Because of the low cost and lower environmental impact, they are also often considered expendable. This is important when measuring in regions where the risk of loss and damage is high. More importantly, these instruments are typically one or more orders of magnitude smaller and lighter than traditional wave buoys and can thus measure in locations where traditional buoys cannot. Crucially, they also offer the opportunity to measure quantities that larger buoys are incapable of capturing, such as the acceleration in breaking waves \citep{feddersen2023CrossshoreWindinducedChanges}. The limits and capabilities of the instrument determine which questions can be answered. Heterogeneous and different solutions are valuable in themselves in that they have different biases, cover different use cases and make results more robust.

A relevant example to this work is \citet{brown2018KinematicsStatisticsBreaking} who previously quantified breakers in the open ocean by measuring accelerations. Similarly, the microSWIFT \citep{thomson2012WaveBreakingDissipation,thomson2019NewVersionSWIFT} has been used to study waves near the shore and in areas with strong currents \citep{rainville2022MeasurementsNearshoreWaves,rainville2023MeasurementsNearshoreOceansurface}.  \citet{sinclair2014FlowRiderLagrangianFloat} and later \citet{brown2021AccelerationsWaveMeasurement},  \citet{fisher2021RapidDeterministicWave}, and \citet{feddersen2023WavedrifterLowcostIMUbased} have attempted to measure breaking waves in the surf or in wave laboratories.


In this study we describe and demonstrate a small, lightweight, buoy, dubbed the Small Friendly Buoy (SFY), tailored to coastal waters and designed for capturing and quantifying breaking processes as well as spatially varying wave-current interactions. The code and hardware is open source and available publicly\footnote{\url{https://github.com/gauteh/sfy}}.
The work builds on the OpenMetBuoy \citep{rabault2020OpenSourceVersatile}.

Some  open-source buoys have telemetry so that the data can be transmitted to shore. This is typically done using Iridium satellites and the Iridium Short Burst Data (SBD) system, which admit only very low bandwidth (except with high-power, heavier, equipment). Some buoy designs can only store the information on board and thus require the buoy to be retrieved. This is often not possible or requires much time and resources. Loss of data is also an issue with such an approach, especially for deployments in rough conditions. The SFY avoids both these problems by relying on the cellular network to transmit data.

The main improvements of the SFY buoy is that it allows us to transmit much more data, is lightweight, and low cost with a small physical and environmental footprint. This combination provides a number of advantages: It can transmit continuous high-frequency time series of the full three-dimensional acceleration vector measured by the buoy. It can be deployed easily from surf-boards or small vessels, or in lightweight moorings. It can be deployed in arrays of many buoys capturing the spatial variability, which is perhaps more important than a single high-quality point measurement in coastal environments. And it can resolve the phase, impulses and trajectory of waves and breaking waves, also between buoys in the same dense spatial network. The buoy does not currently have a magnetometer, but tracks the orientation in its own reference system on short time scales. The deployments must be within the cellular coverage and will not work far from the coast unless there is a marine network available.


\subsection*{Measuring the wave field through acceleration at a point}
\label{subsec:imu-theory}

The working principle of the SFY wave buoy is to measure the acceleration in the wave field, and to convert acceleration as a function of time to displacement as a function of time. This is an established method that has been used for many decades (see e.g.\ \citet{tucker1958AccuracyWaveMeasurementsb}), initially with mechanical accelerometers, and more recently with increasingly cheap, accurate and easily available micro-electromechanical system (MEMS) accelerometers and gyroscopes on inertial measurement units (IMU). Since acceleration ($a$) is the second derivative of displacement, we can recover the displacement ($\eta$) by integrating the acceleration twice,
\begin{subequations}
\begin{align}
  v(t) = v_0 + \int_0^t a(t') \mathrm{d} t', \label{eq:int-v} \\
  \eta(t) = z_0 + \int_0^t v(t') \mathrm{d} t'. \label{eq:int-eta}
\end{align}
\end{subequations}
In practice, some signal processing is required to get a reliable signal, as the measured acceleration typically contains low-frequency noise that would be magnified by the double integration. The common practice is to use some variant of a high-pass filter to remove lower frequencies than those we expect to be present in the wave field (lower than around $0.03 \textrm{--} 0.10$~Hz, corresponding to wave periods in the range $10 \textrm{--} 30$ seconds, see, e.g., \citet{tucker1958AccuracyWaveMeasurementsb, waseda2018CorrelatedIncreaseHigh}). When analyzing the elevation time-series we detrend the signal and apply a Butterworth-filter of 10th order before each integration (described in more detail in Sec.~\ref{sec:lab-experiment}\ref{subsec:lab-significant-wave-height} and \ref{sec:discussion}\ref{subsec:low-frequency-noise}). The integration is performed in the frequency domain (discarding $\omega = 0$),
\begin{equation}
  \mathcal{F}\{\eta(t)\}(\omega) = -\omega^{-2}\; \mathcal{F}\{a(t)\} \label{eq:elevation-freq}.
\end{equation}
Here, $\omega = 2\pi f$ is the angular frequency [rad/s], $f$ is the linear frequency [Hz] and $\mathcal{F}$ is the Fourier transform.
When analyzing spectra of elevation energy (variance density spectrum), $E$, no detrending and filtering is necessary. The integration of the acceleration density ($A$, found using Welch's method), Eq. \eqref{eq:elevation-freq}, is performed in one step:
\begin{equation}
  E(\omega) = \omega^{-4} A(\omega) \label{eq:elevation-energy-freq},
\end{equation}


Assuming the sea-surface elevation is a stationary, random Gaussian process, the characteristics of the variance density spectrum is defined by its moments $m_n = \int_0^\infty f^n E(f) df$. And the significant wave height in deep-water can be estimated as (see e.g. \citet{holthuijsen2010WavesOceanicCoastal} Chapters 3.5 and 4 for a thorough explanation):
\begin{equation}
    H_{m0} \simeq 4 \sqrt{m_0}. \label{eq:hm0}
\end{equation}

There are inherent limitations to measuring a single point of the surface, which cannot capture the changing interference patterns produced by the dispersive wave field \citep{magnusson1999EstimatingExtremesEvolving}. These are better resolved with methods that instantaneously span the spatial field, like stereo-video measurements (see, e.g., \citet{benetazzo2006MeasurementsShortWater,malila2022GroupinessIntermittencyOceanic,malilamika2022InvestigationDynamicalStatistical}) or densely sampled point measurements (e.g \cite{donelan1985DirectionalSpectraWindgenerated,donelan1996NonstationaryAnalysisDirectionala}). Additionally, the motion of wave buoys tend to underestimate the wave crests \citep{longuet-higgins1986EulerianLagrangianAspects,forristall2000WaveCrestDistributions}; moored buoys are affected by the pull of the anchoring and have a tendency to skirt around the highest wave crests, while drifting buoys overestimate the mean water level by spending too much time near the crests.

An inherent assumption of the accelerometer-based approach is that the acceleration experienced by the buoy is the same as the acceleration of the water surface. This assumption starts to break down when the wave oscillations are so rapid that the buoy can no longer reasonably be said to remain at its equilibrium level of immersion. The frequency at which the buoy no longer manages to follow the sea-surface is a function of its shape and buoyancy and can be approximated using the heave response frequency, $f_\mathrm{hr}$  \citep{thomson2015BiofoulingEffectsResponse},
\begin{align}
    \label{eq:f-heave-response}
    f_\mathrm{hr} &= \frac{1}{2\pi} \sqrt{\frac{g \rho V}{ML}}.
\end{align}
Here $\rho$ is the density of seawater, $V$ is the displaced volume, $L$ is the height of the buoy, and $M$ is the mass of the buoy and the added mass caused by drag on the water.

The SFY buoy has been deployed in a small free-drifting form and in a larger moored form. The free-drifting buoy has a heave response of approximately $f_{\mathrm{hr}} = 3.56$ Hz, while the moored buoy has $f_\mathrm{hr} = 1.44$ Hz (assuming the buoy is 50 \% submerged). Clearly, though, the expression for the heave response is an approximation, and does not take the horizontal length of the buoy into account. At a rule-of-thumb level, one can argue that if the wavelength ($\lambda$) of the wave to be measured is shorter than twice the (horizontal) radius of the buoy, $r$, the buoy will not be able to follow the wave. In deep water, the linear phase speed is $c_\mathrm{p} = \sqrt{g/k}$ with $k=2\pi/\lambda$ the wavenumber. From this we find that the geometric cut-off frequency ($f_{\mathrm{gc}}$) where the wavelength equals $2r$ is
\begin{align}
    \label{eq:f_max}
    f_{\mathrm{gc}} = \frac{\sqrt{g/k}}{\lambda} = \sqrt{\frac{g}{4 \pi r}}. 
\end{align}
For the free-drifting buoy $f_{\mathrm{gc}} = 2.24$ Hz, and for the moored buoy $f_{\mathrm{gc}} = 1.09$ Hz. In shallower waters a different phase speed should be used to get a more accurate cut-off frequency. These frequencies will be compared with the spectra from experiments in Sections~\ref{sec:lab-experiment} and~\ref{sec:fedjeosen}.

\subsection*{Structure of this paper}

This paper is organised as follows. We first present the design of the SFY buoy in Section~\ref{sec:sfy-design}. Wave tank experiments of the buoy response are presented in Section~\ref{sec:lab-experiment}. Field experiments from a mooring in open-ocean conditions are presented in Section~\ref{sec:fedjeosen} together with a comparison against a high-resolution wave model. Results from surf-zone deployments are presented in Section~\ref{sec:surf-experiments}. Finally, we discuss our findings and compare with other light-weight buoys in Section~\ref{sec:discussion}, and present our conclusions in Section~\ref{sec:conclusions}.

\section{Buoy design}
\label{sec:sfy-design}
The SFY wave buoy consists mainly of three parts; (i) the electronics board with sensors, processing and telemetry (presented in Section~\ref{subsec:electronics}); (ii) the power source, usually alkaline or lithium batteries (see appendix~\ref{app:power}); (iii) the housing which protects the other parts and keeps the buoy floating. Various types of housing can be used, but in this work two types are described, an 800 ml bottle for free-drifting buoys, and a $60\times 80$ cm inflatable fender for moored buoys.

\subsection*{Sensor, processing and telemetry}
\label{subsec:electronics}
\begin{figure}[h]
  \centering
  \includegraphics[width=\hsize]{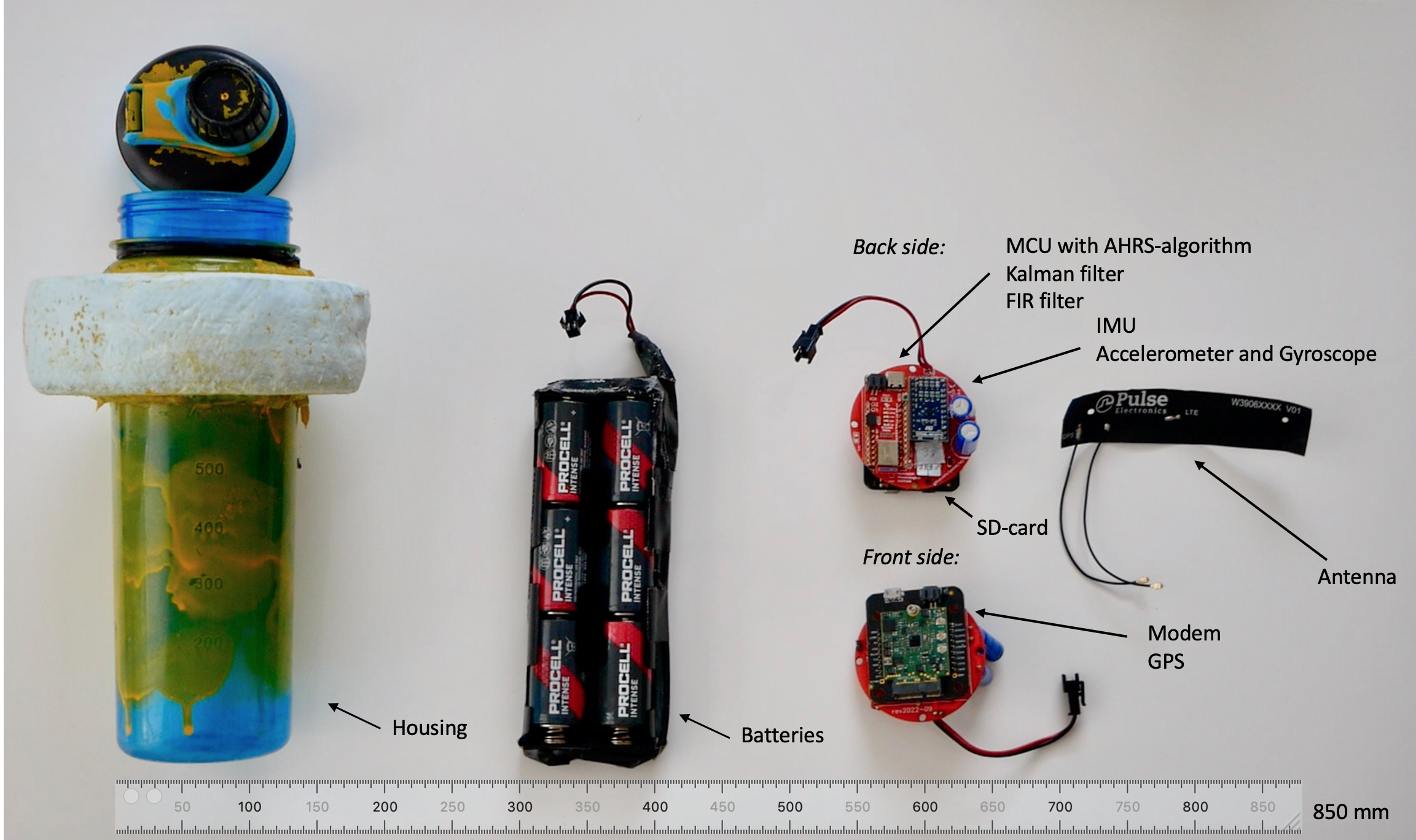}
  \caption{The components that make up the SFY wave buoy. From the left: housing (plastic bottle) with foam for additional stability, battery pack (alkaline), main board shown twice (both sides): Microcontroller Unit (MCU) in red, modem and GPS in green, SD-card, IMU, and GPS and cellular antennas. For scale: the ruler is 85 cm long.}
  \label{fig:components-buoy}
\end{figure}

The circular electronics board measures 60 mm in diameter and about 2 cm in height. Figure~\ref{fig:components-buoy} shows the components with the 800 ml bottle used for housing. Some additional weight is added in order to position the bottle upright in the water. The electronics are placed in bags of bubble wrap, and the antennas positioned as high and unobstructed as possible near the cap. The cap is sealed with self-amalgamating tape. This setup is robust against high impacts and keeps the water out.

The sensor pipeline generally follows the setup for the OpenMetBuoy \citep{rabault2022OpenMetBuoyv2021EasytoBuildAffordable}, hereafter OMB, and benefits from the extensive field testing and validation that the OMB has been subjected to. The inertial measurement unit (IMU) used is either the ISM330DHCX or LSM6DSOX \citep{stmicroelectronics2019LSM6DSOXDatasheet,stmicroelectronics2020ISM330DHCXDatasheet}, both of which yield gyroscope and accelerometer readings. The IMU low-pass filters the signal
internally according to the configured output rate. The theoretical noise level of the ISM330DHCX is typically $60\,\mu g / \mathrm{Hz}^{1/2}$ (maximum $100\, \mu g / \mathrm{Hz}^{1/2}$), according to the manufacturers data sheet \citep{stmicroelectronics2020ISM330DHCXDatasheet}. Here, $g = 9.81\,\mathrm{m\,s^{-2}}$ is the earth's gravitational acceleration.

Accelerometer and gyroscope readings from the IMU are fused by applying an Attitude and Heading reference system (AHRS) algorithm implemented with a Kalman filter so that the acceleration is obtained in a North-East-Down (NED) frame of reference \citep{nxpsemiconductors2016NXPSensorFusion,adafruit2023AdafruitAHRSLibrary}. No magnetometer is currently included, so the North-East vector points in an arbitrary direction, but remains stable on short time scales thanks to the measurements provided by the gyroscope. The horizontal direction is therefore only analyzed on short timescales in the buoy reference system (See Sec. \ref{sec:sfy-design}\ref{subsec:lab-traj}). The IMU is sampled at $208$ Hz and the samples are then fed into the Kalman filter at this same frequency. Finally, the output from the Kalman filter is filtered using a finite-impulse response (FIR) low-pass filter (129 taps with a Hamming window) and decimated down to $52$ Hz or $26$ Hz depending on the application.


It is difficult to quantify the theoretical noise level of the processing chain as a whole since both the accelerometer and gyroscope introduce noise before these are fused in a complex nonlinear way by the Kalman filter. Hence, the noise level of the accelerometer is used as a lower bound of the noise and as an estimate of its frequency dependence. The noise level of the sensors and the processing pipeline as a whole are discussed in more detail by \citet{rabault2023DatasetDirectObservations}. Details on how the data are packed, transmitted and received are described in appendix~\ref{app:processing}.

\section{Lab experiment: Wave flume}
\label{sec:lab-experiment}
Experiments were carried out in a wave flume at SINTEF Ocean in Trondheim, Norway (described in \cite{singsaas2020FateBehaviourWeathered}). The flume is 12 m long and 0.53 m wide, with a piston-type wave maker from Edinburgh Designs \citep{edinburghdesigns2023PistonCoastalWave} at one end, and a perforated absorbing beach at the other end. The wave maker is programmable and can be set to generate simple harmonic waves as well as wave packets and random realisations of a specified wave spectrum. In the experiment below it is configured to generate harmonic waves of different amplitudes and frequencies, as well as focused breaking waves.

The flume is filled with filtered seawater from an intake at approximately 80 m depth in the Trondheim fjord. During the experiments described here, the flume was filled to a depth of 80 cm. Wave gauges can be mounted at any position in the flume, allowing the surface elevation to be sampled at frequencies up to 128 Hz. The wave gauges, also produced by Edinburgh Designs, are based on the principle of measuring resistance between two vertical parallel metal rods. Table~\ref{tab:lab-buoy-setup} describes the configuration of the five buoys and wave gauge used.

\begin{table}[h]
  \centering

  \begin{tabular*}{\hsize}{ c c c c l c }
    \hline\hline
    Instrument  &           & Frequency &         & IMU      \\
                & IMU       & Output    & Real    &          \\
    \hline
    Wavebug16   &   52 Hz   & 52 Hz  & 48.58 Hz & LSM6DSOX   \\
    Wavebug23   &   52 Hz   & 52 Hz  & 48.60 Hz & LSM6DSOX   \\ 
    Wavebug25   &   208 Hz  & 52 Hz  & 49.07 Hz & ISM330DHCX \\
    Wobbler001  &   52 Hz   & 52 Hz  & 55.90 Hz & ISM330DHCX \\
    Wobbler002  &   52 Hz   & 52 Hz  & 55.80 Hz & ISM330DHCX \\
    Wave gauge  &           & 128 Hz & 128.00 Hz   &         \\
    \hline\hline
  \end{tabular*}

  \caption{Buoy and wave gauge configuration for lab experiment. See appendix~\ref{app:processing} for more information on IMU frequency discrepancies.}
  \label{tab:lab-buoy-setup}

\end{table}

\begin{figure*}[t]




  \includegraphics[width=\textwidth]{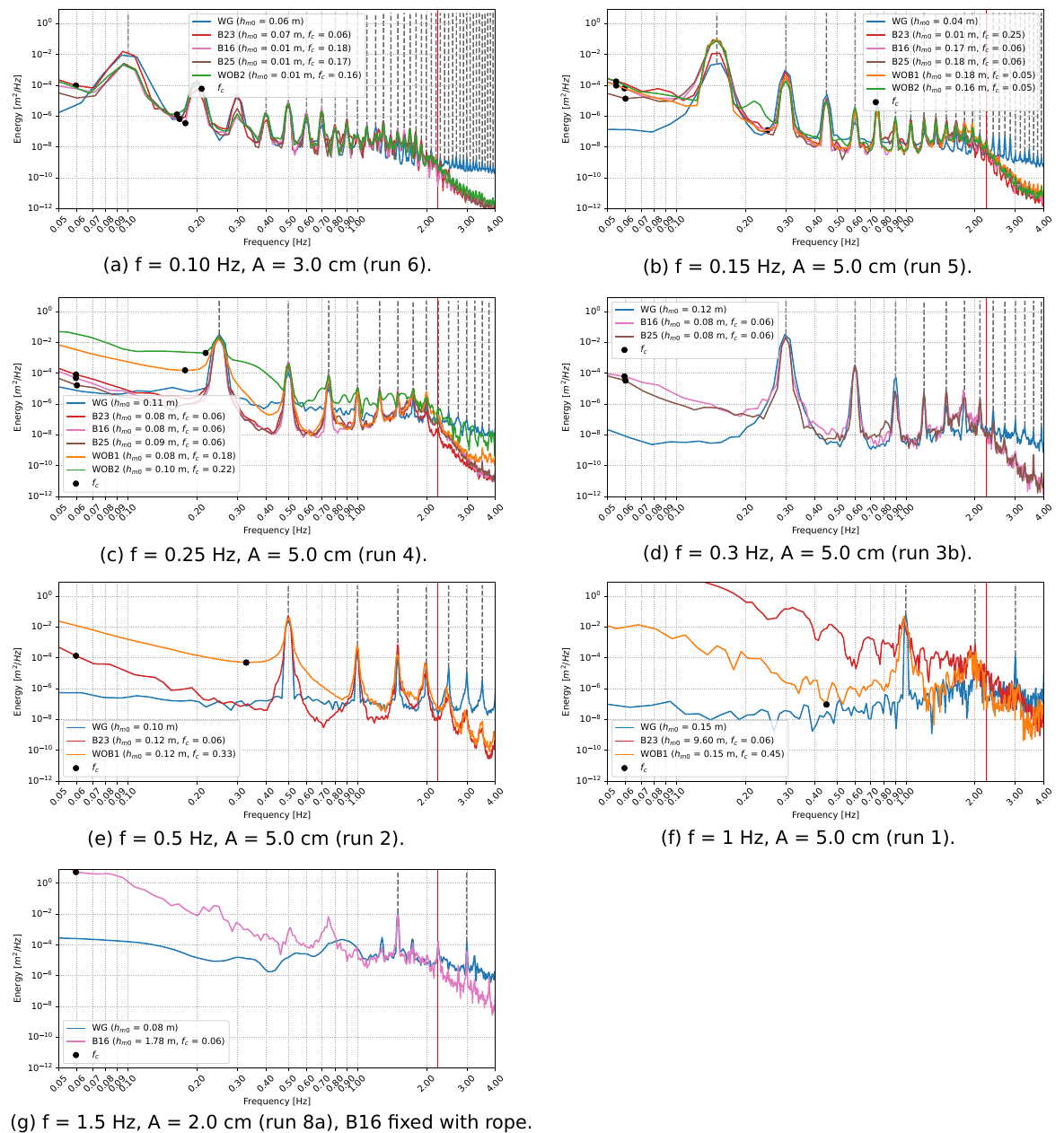}
  \caption{Welch spectrum measured by buoys and wave gauge (WG) measured in a wave flume with piston configured at different frequencies and wave amplitudes (shown for $0.05\textrm{--}4$ Hz). The red vertical line marks frequency where the wavelength is twice the size of the collar width of the buoys ($2.24$ Hz, see Eq.~\eqref{eq:f_max}). $f_c$ (black dot) marks the cut-off frequency for the low-frequency noise common to acceleration based measurements, discussed below and extensively in Sec.~\ref{sec:discussion}\ref{subsec:low-frequency-noise}. The wave gauge measures elevation directly and is less vulnerable to the low-frequency noise since it is not integrated (Eq. \eqref{eq:int-eta}). The vertical dashed lines mark the input frequency and its higher harmonics.}
  \label{fig:lab-welch-spectrums}
\end{figure*}

The Welch spectrum in Fig.~\ref{fig:lab-welch-spectrums} is calculated with a segment length of 4096 for the buoys, and 8192 for the wave gauge which samples at approximately twice the frequency of the wave-buoy, both are detrended using a linear least square fit\footnote{\url{https://docs.scipy.org/doc/scipy/reference/generated/scipy.signal.welch.html}}.

The measured peaks in the spectra (Fig.~\ref{fig:lab-welch-spectrums}) match the configured piston frequency and its higher harmonics (dashed gray vertical lines) well. The frequency of the wave components measured with the wave gauge also match very well. In the initial experiments the buoys were freely moving at different positions in the wave flume and were subjected to considerable lateral movement and rotation due to Stokes drift, so the phase between the buoys should not be expected to match. Additionally, the vertical amplitude of the generated wave varied considerably laterally along the flume, probably due to interference and reflection (particularly for the lower frequencies).

At approximately $2.2$ Hz there is a marked decrease in the wave-buoy spectra compared to the wave gauge. The collar-width of the buoy (Fig.~\ref{fig:components-buoy}) is about $15.5$ cm. From Eq.~\eqref{eq:f_max}, we see that a wavelength twice that has a deep-water frequency of $2.24$ Hz (red vertical line in Fig.~\ref{fig:lab-welch-spectrums}). The measured amplitude is therefore increasingly dampened as the wavelength approaches and becomes shorter than the diameter of the buoy. According to Eq.~\eqref{eq:f-heave-response}, the heave response frequency is $f_\mathrm{hr} = 3.56$ Hz, but it is very sensitive to the degree of immersion and displacement volume. It seems that the dominating effect on the heave response is the diameter of the buoy, rather than the weight and volume. \citet{yurovsky2020MEMSbasedWaveBuoya} observe a similar geometric cut-off frequency of about $2.5$ Hz (corresponding to the wavelength twice the hull diameter of their buoy). As they mention, this means that shorter waves can be measured compared to most other alternatives (e.g., $1.2$ Hz for the Spotter buoy \citep{raghukumar2019PerformanceCharacteristicsSpotter}).

The buoys measure acceleration and have an approximately constant noise level on the measured acceleration. The wave gauge measures displacement and has an approximately constant noise level on the displacement. The difference becomes apparent in the spectrum (Fig.~\ref{fig:lab-welch-spectrums}) at lower frequencies, where the noise level of the wave gauge is flat (horizontal), while the noise level of the wave buoys is proportional to $f^{-2}$ following the double integration from acceleration to displacement. At lower frequencies the wave buoys therefore have a higher noise ratio compared to the wave gauge, whereas at higher frequencies the wave buoy performs better since acceleration is better for measuring high frequency waves (with higher acceleration than displacement). See Sec.~\ref{sec:introduction} and~\ref{sec:discussion}\ref{subsec:low-frequency-noise} for more details.

\subsection{Significant wave height}
\label{subsec:lab-significant-wave-height}

\begin{table*}[t]
  \centering

  \begin{tabular}{ c c c | c c l c c c}
    \hline\hline
    \multicolumn{3}{l|}{Piston} & \multicolumn{6}{|c}{$H_{m0}$ (m)}  \\
    \hline
    $f_p$ (Hz) & Run & $A_p$ / $H_{m0}$ (m) & WG   & B16  & B23  & B25  & WOB1 & WOB2 \\
    \hline
    0.10 & 6  & 0.03 / 0.08 & 0.06 & 0.01 & 0.07 & 0.01 &      & 0.01 \\ 
    0.15 & 5  & 0.05 / 0.14 & 0.04 & 0.17 & 0.01 & 0.18 & 0.18 & 0.16 \\ 
    0.25 & 4  & 0.05 / 0.14 & 0.11 & 0.08 & 0.08 & 0.09 & 0.08 & 0.10 \\ 
    0.30 & 3a & 0.05 / 0.14 & 0.12 &      & 0.13 &      & 0.09 & \\ 
    0.30 & 3b & 0.05 / 0.14 & 0.12 & 0.08 &      & 0.08 &      & \\ 
    0.50 & 2  & 0.05 / 0.14 & 0.10 & 0.12 &      & 0.12 &      & \\ 
    1.00 & 1  & 0.05 / 0.14 & 0.15 &      & $9.60^\dagger$ &  &      0.15 & \\ 
    1.50 & 7a & 0.03 / 0.08 & 0.08 & $4.68^\dagger$ & $7.11^\dagger$ & $4.23^\dagger$ &      & $2.14^\dagger$ \\ 
    1.50 & 7b & 0.03 / 0.08 & 0.08 & $3.43^\dagger$ &      &      &      & \\ 
    1.50 & 8a & 0.02 / 0.06 & 0.08 & $1.78^\dagger$ &      &      &      & \\ 
    \hline\hline
  \end{tabular}

  \caption{Significant wave height measured in the wave flume, automatic high-pass cut-off frequency. Values marked with $\dagger$ are unreasonably large because the automatic cut-off frequency is chosen to low, and includes low-frequency noise. }
  \label{tab:lab-buoy-hm0}

\end{table*}

In Table~\ref{tab:lab-buoy-hm0} the significant wave height is calculated from the zeroth moment of the elevation variance spectrum ($H_{m0}$, Eq.~\eqref{eq:hm0}) and compared to the programmed wave amplitude of the piston, and the significant wave height ($H_{m0}$) of the wave gauge (WG). The theoretical $H_{m0}$ for the generated signal with a single harmonic wave with amplitude $A_p$ and variance $\frac{1}{2}A_p^2$ is calculated as (see \citet{holthuijsen2010WavesOceanicCoastal}, Chapter 3.5):
\begin{equation}
  H_{m0} = 4 \sqrt{m_0} = 4 \sqrt{\frac{1}{2}A_p^2} \label{eq:hm0-harmonic}
\end{equation}
The measured signal from the buoys need to be high-pass filtered at a cut-off frequency, $f_c$, to avoid low-frequency noise. The wave gauge is not filtered, since it has a flat noise curve on displacement (Fig.~\ref{fig:lab-welch-spectrums}). Automatic selection of low-end cut-off frequency is generally hard. This is a problem that affects IMU based buoys in general (see, e.g., \citet{nose2018PredictabilityStormWavea}) and will be discussed further in Section~\ref{sec:discussion}\ref{subsec:low-frequency-noise}. \citet{alari2022LainePoissLightweightIceResistant} chose the bandpass cut-off frequencies of the measured signal in the wave tank tests to be $0.30$ and $1.28$ Hz and the \emph{LainePoiss\textsuperscript{®}} buoy was evaluated at frequencies between $0.57$ and $0.73$ Hz.

In Table~\ref{tab:lab-buoy-hm0} the cut-off frequency is found automatically by searching for the first spectral minimum above $0.05$ Hz using the \emph{find\_peaks} algorithm in SciPy\footnote{\url{https://github.com/gauteh/sfy/blob/v1.0/sfy-processing/sfy/signal.py#L319}} (minimum distance of $7$ frequency bins for $52$ Hz sample rate, $3$ for $26$ Hz sample rate, and a prominence of $0.05$). The algorithm is based on the process outlined in \citet{rabault2022OpenMetBuoyv2021EasytoBuildAffordable} (Figure 7) as well as in \citet{tucker1958AccuracyWaveMeasurementsb, waseda2018CorrelatedIncreaseHigh}. However, when this algorithm is unsuccessful it results in unphysically large amplitudes and significant wave height estimates. The auto-detected cut-off frequency is shown and listed in Fig.~\ref{fig:lab-welch-spectrums}. For instance for $1$ Hz in Table~\ref{tab:lab-buoy-hm0} and Fig.~\ref{fig:lab-welch-spectrums}f the significant wave height for \emph{B23} is obviously wrong and overestimated compared to \emph{WOB1}. This can be attributed to the cut-off frequency being chosen too low for \emph{B23}, while better for \emph{WOB1} (see $f_c$ in Fig.~\ref{fig:lab-welch-spectrums}f, note that the cut-off for \emph{B23} is off the scale of the plot). In general it is easier to auto-detect the cut-off frequency when the waves are higher and the signal to noise ratio is greater. Since we know the input frequency for these experiments we can manually choose cut-off frequencies. In Table~\ref{tab:lab-buoy-hm0-bandpassed} the significant wave height is shown for the manually chosen frequency band. These values match better, and do not show exaggerated significant wave heights like those present in Table~\ref{tab:lab-buoy-hm0} (where a too low or too high cut-off frequency has been picked). In unknown conditions it is therefore best to start with a minimum of $0.05$ Hz for this buoy, and possibly other IMU based buoys. If the wave height suddenly jumps to unrealistic levels or there is significant energy at very low frequencies this is a sign that it should be increased, or that the measurements have been compromised in other ways (e.g. a poor mooring, unstable float or interference from traffic).

\begin{table*}[t]
  \centering

  \begin{tabular}{ c c c c c | c c l c c c}
    \hline\hline
    \multicolumn{3}{c}{Piston} & \multicolumn{2}{c}{Bandpass} & \multicolumn{6}{c}{$H_{m0}$ (m)}  \\
    \hline
    $f_p$ (Hz)                                     & Run & $A_p$ / $H_{m0}$ (m) & $f_{hp}$ & $f_{lp}$ & WG   & B16  & B23  & B25  & WOB1 & WOB2 \\
    \hline
    0.10                                           & 6   & 0.03 / 0.08 & 0.05     & 1.28     & 0.06 & 0.03 & 0.07 & 0.03 &      & 0.03 \\ 
    0.15                                           & 5   & 0.05 / 0.14 & 0.05     & 1.28     & 0.04 & 0.17 & 0.07 & 0.18 & 0.18 & 0.16 \\ 
    0.25                                           & 4   & 0.05 / 0.14 & 0.10     & 1.50     & 0.11 & 0.08 & 0.08 & 0.09 & 0.08 & 0.12 \\ 
    0.30                                           & 3a  & 0.05 / 0.14 & 0.10     & 1.50     & 0.12 &      & 0.13 &      & 0.09 & \\ 
    0.30                                           & 3b  & 0.05 / 0.14 & 0.10     & 1.50     & 0.12 & 0.08 &      & 0.08 &      & \\ 
    0.50                                           & 2   & 0.05 / 0.14 & 0.30     & 2.00     & 0.10 &      & 0.12 &      & 0.12 & \\ 
    1.00                                           & 1   & 0.05 / 0.14 & 0.30     & 2.00     & 0.15 &      & 0.43 &      & 0.15 & \\ 
    1.50                                           & 7a  & 0.03 / 0.08 & 0.30     & 2.00     & 0.08 & 0.18 & 0.23 & 0.15 &      & 0.08 \\ 
    1.50                                           & 7b  & 0.03 / 0.08 & 0.30     & 2.00     & 0.08 & 0.11 &      &      &      & \\ 
    1.50                                           & 8a  & 0.02 / 0.06 & 0.30     & 2.00     & 0.08 & 0.10 &      &      &      & \\ 

    \hline\hline

  \end{tabular}

  \caption{Significant wave height measured in the wave flume, custom bandpass cut-off frequencies.}
  \label{tab:lab-buoy-hm0-bandpassed}
\end{table*}

\subsection{Wave buoy orbital measurements}
\label{subsec:lab-traj}

\begin{figure}[h]
  \centering
  \includegraphics[width=\hsize]{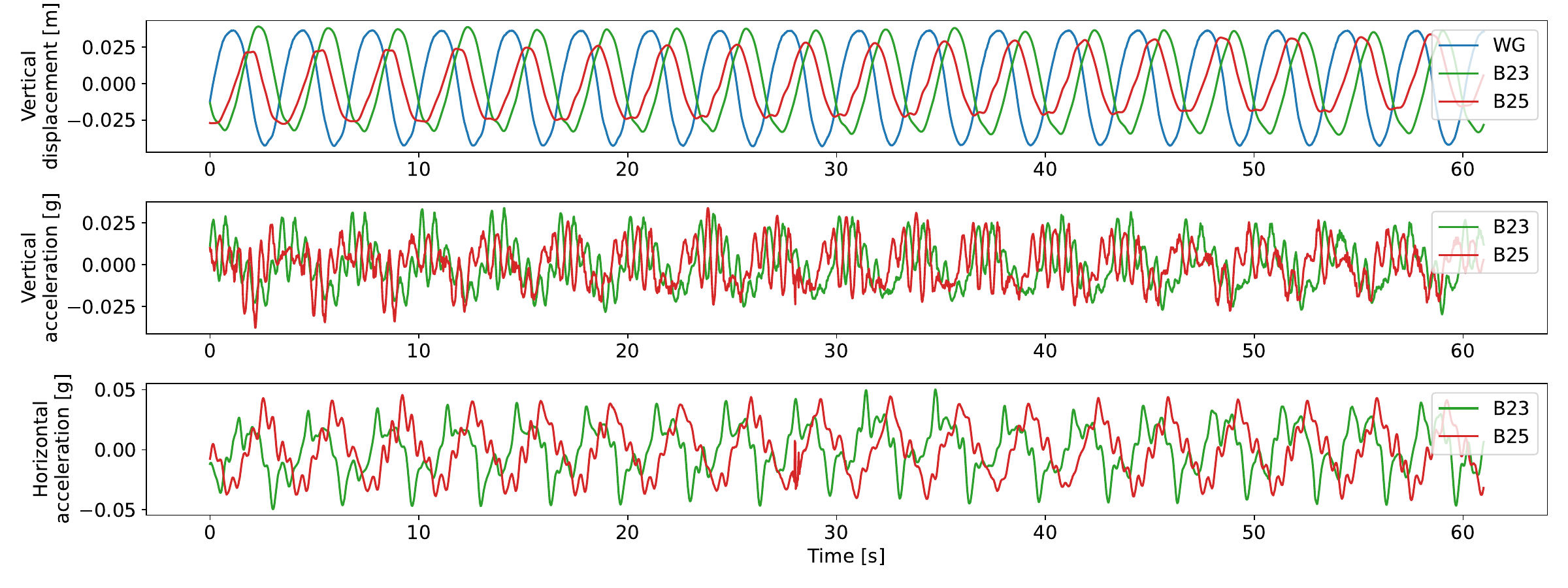}
  \caption{Vertical motion, vertical acceleration and horizontal acceleration along direction of maximum variance for a $0.3$ Hz sine wave. \emph{Upper:} Vertical displacement for the wave-gauge, B23 and B25. \emph{Middle:} Vertical acceleration for B23 and B25. \emph{Lower:} Horizontal acceleration for B23 and B25.}
  \label{fig:lab-traj-uz}
\end{figure}

To determine the buoy orientation along the wave-flume we assume that the buoy movement is mainly varying along the wave direction. Since we do not have a magnetometer we do not expect the $x$ and $y$ axes to be pointing in the same direction on long time-scales (limited by gyroscope drift). Figure~\ref{fig:lab-traj-uz} shows the vertical motion as a function of time for a sine wave of $0.3$ Hz (Fig.~\ref{fig:lab-welch-spectrums}d), as well as the vertical and horizontal acceleration.

In order to obtain the trajectory of the wave buoy we apply principal component analysis (PCA) to the high-pass filtered $x$ and $y$ acceleration. The acceleration is filtered (using the Butterworth-filter) so that only movement due to wave motion is included. The first principal component, $\bf u_0$, is the direction in the $xy$-plane along which the variation is greatest, the second component ($\bf u_1$) is orthogonal to $\bf u_0$ representing the rest of the variance.

\begin{figure}[h]
  \centering
  \includegraphics[width=\hsize]{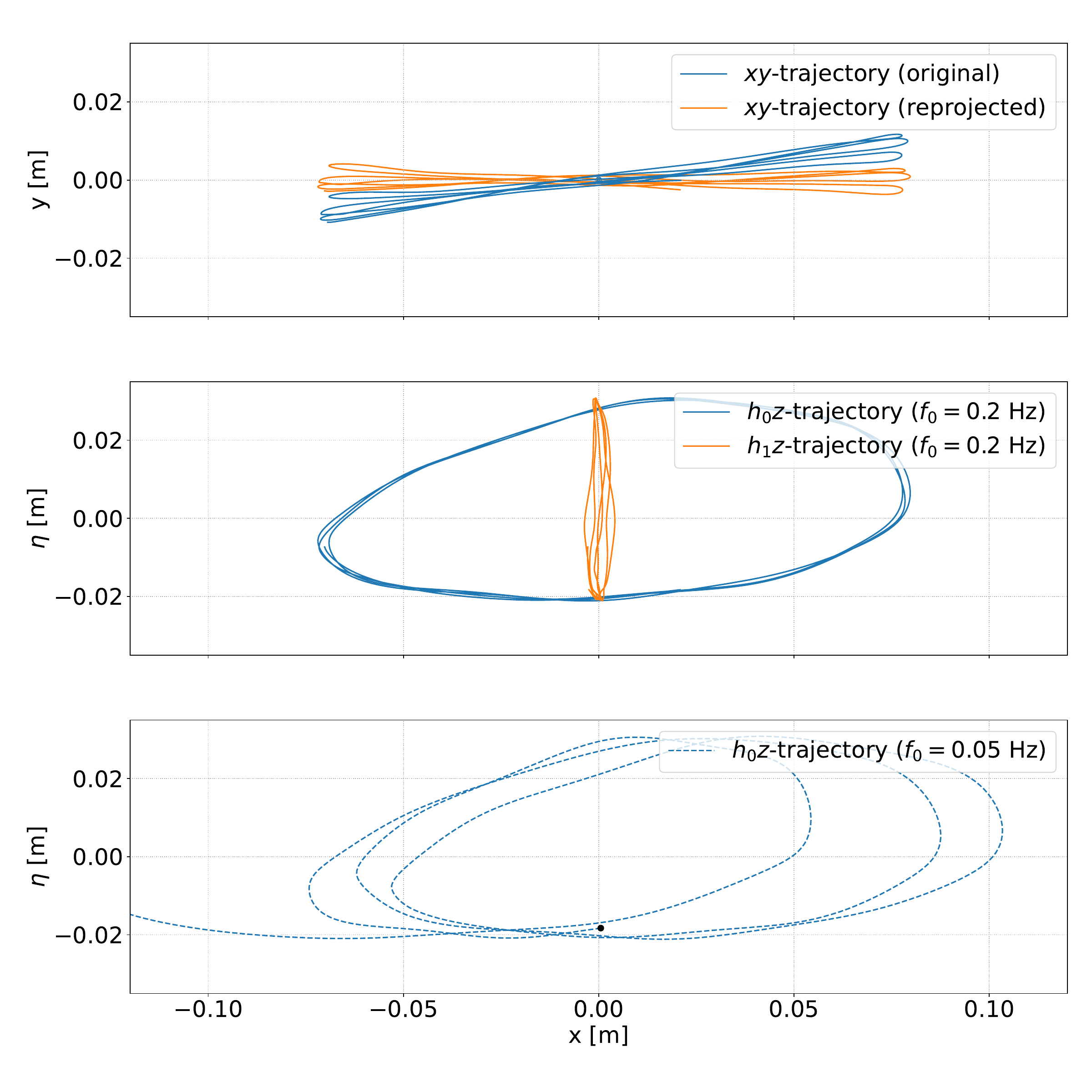}
  \caption{Trajectory of the \emph{B25} wave-buoy in the wave-flume with about 3 cycles of a $0.3$ Hz (10 seconds) sinusoidal wave (Fig.~\ref{fig:lab-welch-spectrums}d). $f_0$ is the lower cut-off frequency in the bandpass filter when integrating the acceleration and computing the principal components. \emph{Upper:} \emph{xy}-plane with principal components and re-projected trajectory. \emph{Middle:} high-pass filtered trajectory in horizontal--vertical plane along first ($\bf u_0$) and second ($\bf u_1$) principal component. \emph{Lower:} full bandwidth trajectory in horizontal--vertical plane for first principal component.}
  \label{fig:lab-traj-hz}
\end{figure}

We project the $x$ and $y$-accelerations onto the normalized first principal component ($\frac{\bf u_0}{\left\| \bf u_0 \right\|}$, horizontal) and the normalized second principal component ($\frac{\bf u_1}{\left\| \bf u_1 \right\|}$, vertical), the original movement in the $xy$-plane and the re-projected movement is shown in Fig.~\ref{fig:lab-traj-hz}:
\begin{equation}
  \begin{bmatrix}
    h_i^1 \\
  \vdots \\
  h_i^N
  \end{bmatrix}
  = \begin{bmatrix}
    x^1 & y^1 \\
    \vdots & \vdots \\
    x^N & y^N
  \end{bmatrix}
    \cdot \frac{\bf u_i}{\left\| \bf u_i \right\|}
\end{equation}
$h_0^n$ and $h_1^n$ are the vector components with re-projected $x$ and $y$ accelerations, and the superscript index $n$ indicates time from $t_1$ to $t_N$.

If the buoy is only moving along one direction, and the gyroscope is not drifting, all movement should be aligned along the first principal component. It is clearly visible in Fig.~\ref{fig:lab-traj-hz} that this is the case for almost all the movement. The ratio between the two principal components (variances) for 10 seconds in the wave-flume is $925$ (upper panel in Fig.~\ref{fig:lab-traj-hz}), which provides a measure of how robust this method is on this time scale.

Linear wave theory predicts the orbital motion of particles. If the wave-buoy is to measure the details of wave-motion it should be water-following. The relation between the horizontal and vertical motion, and their spectra, will tell us how well the buoy is following the particle motion in a wave \citep{feddersen2023WavedrifterLowcostIMUbased}. The amplitudes of the horizontal and vertical particle orbital velocity in a plane wave are (e.g. \citet{holthuijsen2010WavesOceanicCoastal})
\begin{align}
  u_x &= \omega a \frac{\cosh [ k(d + z) ]}{\sinh (kd) } \label{eq:linear-wave-amplitude-x} \\
  u_z &= \omega a \frac{\sinh [ k(d + z) ]}{\sinh (kd) } \label{eq:linear-wave-amplitude-z}.
\end{align}
Here, $d$ the local water depth and $k$ the wave number. The ratio of horizontal to vertical velocity is thus
\begin{equation}
  \frac{u_z}{u_x} = \frac{\sinh [ k(d + z) ]}{\cosh [ k (d + z) ]} = \tanh [ k (d + z) ]. \label{eq:uz_ux_kh}
\end{equation}
At the surface, $z = 0$, Eq. \eqref{eq:uz_ux_kh} becomes $\tanh (kd)$. The same relation \eqref{eq:uz_ux_kh} holds true for the acceleration, $\frac{a_z}{a_x}$ (differentiate Eq. \eqref{eq:linear-wave-amplitude-x} and Eq. \eqref{eq:linear-wave-amplitude-z}). In this setup $k = 0.71\; \textrm{m}^{-1}$ for $f = 0.3$ Hz, and $\tanh (kd)$ is $0.51$, while the orbital motions in Fig.~\ref{fig:lab-traj-hz} (middle panel) have $\frac{u_z}{u_x}$ of approximately $0.36$. We note that for $k=0.71$, the wavelength here is no longer short compared to the length of the flume, and perfect match with theory can not be expected since reflections and standing waves may start to become relevant.

\subsection{Breaking waves}
\label{subsec:lab-breaking-waves}

\begin{figure}[h]
  \centering
  \includegraphics[width=1\hsize]{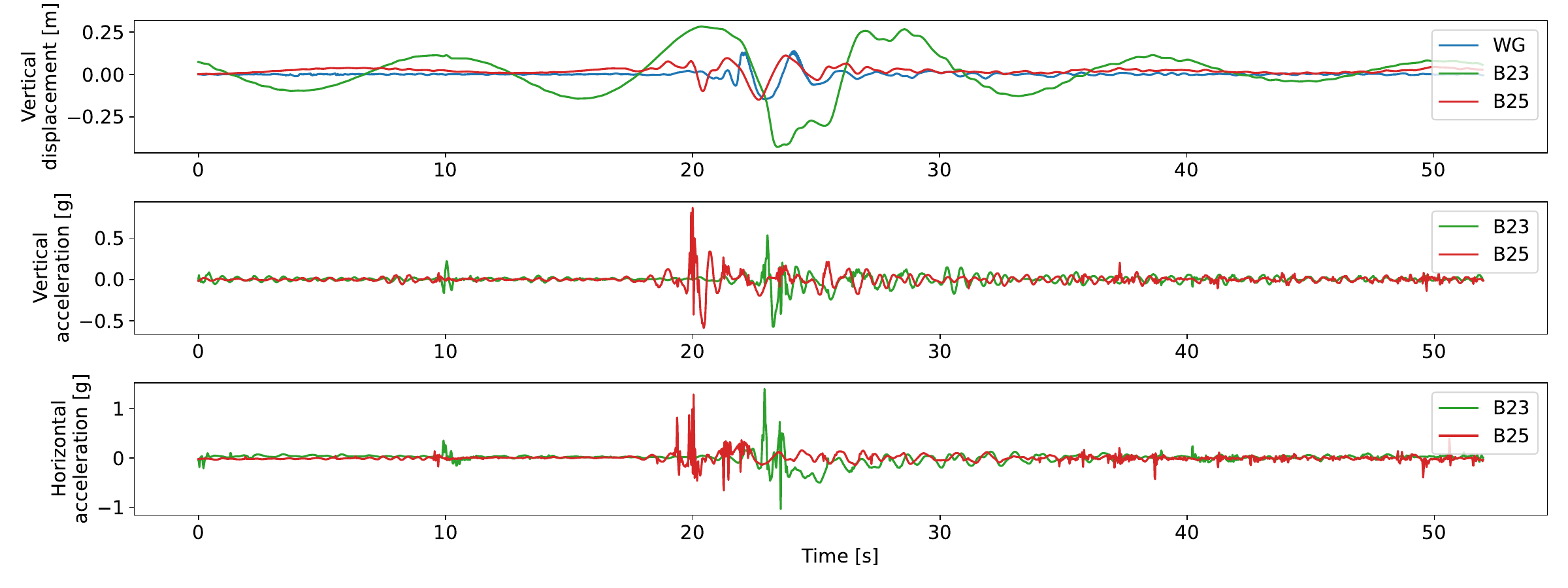}
  \caption{A breaking wave measured in the wave-flume by the buoys drifting freely next to the wave gauge. The integrated vertical displacement for the wave gauge (WG), and buoys 23 and 25 are shown. The vertical and horizontal acceleration (along main principal component, as in Sec.~\ref{subsec:lab-traj}). Buoy 23 samples the IMU at $52$ Hz, while buoy 25 samples the IMU at $208$ Hz.}
  \label{fig:breaker-lab}
\end{figure}


Two observations of the relation between the measured harmonic sine waves and the breaking waves are of particular interest. First, the maximum horizontal acceleration for the sine wave is of the order of $\pm 0.05g$ (Fig.~\ref{fig:lab-traj-uz}), whereas for the breaking wave \emph{in the lab} it exceeds $\pm 1g$. Second, buoys 23 and 25 show similar accelerations and vertical displacements for the non-breaking wave (upper panel in Fig.~\ref{fig:lab-traj-uz}). However, for a breaking wave the integrated vertical displacement (upper panel in Fig.~\ref{fig:breaker-lab}) is dominated by exaggerated low-frequency signals (long period oscillations) for buoy 23. The major difference between the two buoys during the recording of the breaking wave is that buoy 23 is sampling the IMU at $52$ Hz, while buoy 25 is sampling at the full $208$ Hz before filtering and downsampling to $52$ Hz after feeding and re-projecting the acceleration through the AHRS fusion algorithm. These results will be discussed further in Sec.~\ref{sec:discussion}\ref{subsec:low-frequency-noise}.

\section{Open water: Fedjeosen}
\label{sec:fedjeosen}

\begin{figure}[h]
  \includegraphics[width=\hsize]{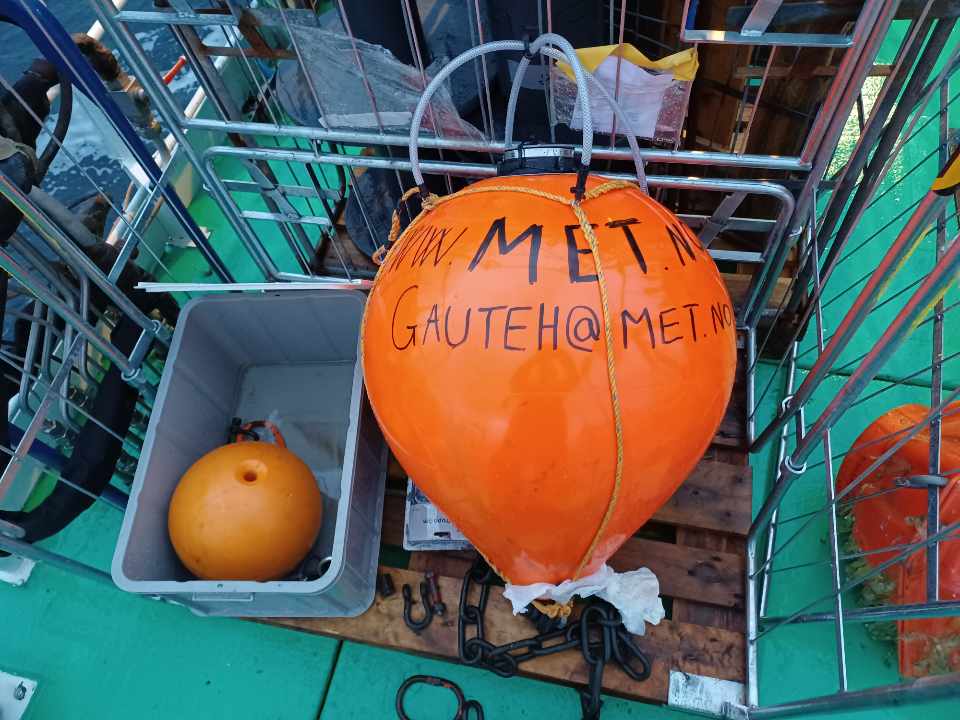}
  \caption{Fedjeosen mooring: A fender of dimensions $65$ cm x $88$ cm floats at the surface, and houses the electronics and batteries.}
  \label{fig:fedjeosen-mooring}
\end{figure}

A wave buoy built into a fender was deployed at $60.752145^\circ$ N, $4.693160^\circ$ E (Figure~\ref{fig:fedjeosen-map}), moored at $90$ m depth, west of Hellisøy lighthouse in Fedjeosen in July 2023. The location is exposed to westerly waves from the North Sea, but is sheltered from the north by Fedje Island. The configuration is described below (shown in Figures~\ref{fig:fedjeosen-mooring} and~\ref{fig:fedjeosen-mooring-schematic}) and is very similar to recommended moorings described by \citet{martini2021MooringDesignOperational}.

Fedjeosen is the main shipping lane into Hjeltefjorden and the coastal installations around Bergen. 
The channel creates a strong tidal current with spatially and temporally varying current gradients across the channel (which is about $2$ km wide). This will give rise to strong wave-current interaction and a spatially inhomogeneous wave field with open-ocean amplitudes. It is thus a challenging location to test the equipment, but also one where both storm waves and low-frequency swell can be found, allowing us to test the full measurement range of the buoy.

The buoy is equipped with $3 \times 21 = 63$ D-cell alkaline batteries, sufficient for a one-year deployment with transmissions over the cellular network of measurements sampled at $26$ Hz.

The buoy was deployed ``anchor last'' with an anchor of approximately $80$ kg. As significant wave heights of more than $12$ m can be expected during winter storms, more weight is required than in sheltered locations. Still, the weight is almost two orders of magnitude less than for large wave buoys, making the deployment manageable with a smaller vessel. The mooring rope is $12$ mm Danline Polypropylene (PP). This diameter dimension is excessive, and causes unnecessary drag from currents, and a smaller diameter should be chosen in the future. The buoy is deployed by first lowering the floating buoy to the water surface, then the mooring is unspooled while traveling against the weather. Finally the anchor can be lifted off the ship and be dropped with a releasable hook. The rope is weighted with a steel wire of approximately $10$ kg between the surface buoy and a buoyancy float at $30$ m so that the buoy is relatively free to move with the waves, and the rope is kept out of the way for vessels traveling around the buoy. The bottom anchor is split in two parts with some chain as buffer, so that in large waves the first chain and anchor will move first in order to prevent the entire anchor from being lifted and wandering off position (see schematic in appendix~\ref{app:fedjeosen-mooring}). The cellular modem connects to a station $80$ km away (more details in appendix~\ref{app:cellular}).

\begin{figure*}[t]
  \centering
  \begin{center}
    \includegraphics[width=\textwidth]{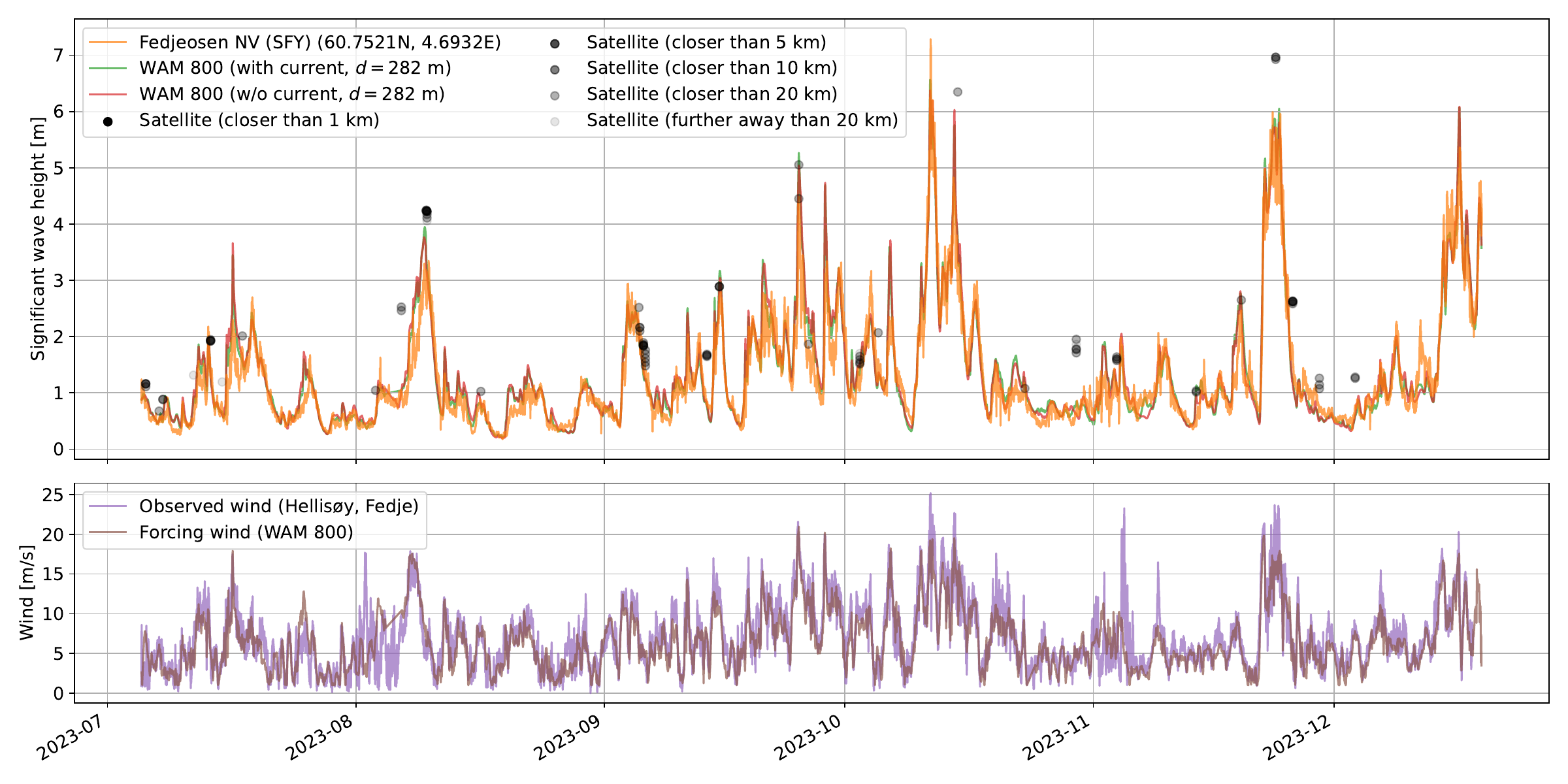}
  \end{center}
  \caption{$H_{m0}$ measured by a moored buoy at the western coast of Norway (Fedjeosen), compared to operational wave models (with and without current) \citep{thewamdigroup1988WAMModelThird}, satellite observations of significant wave height, and the measured wind nearby.}
  \label{fig:fedjeosen_hm0}
\end{figure*}

\begin{figure}[h]
  \centering
  \includegraphics[width=\hsize]{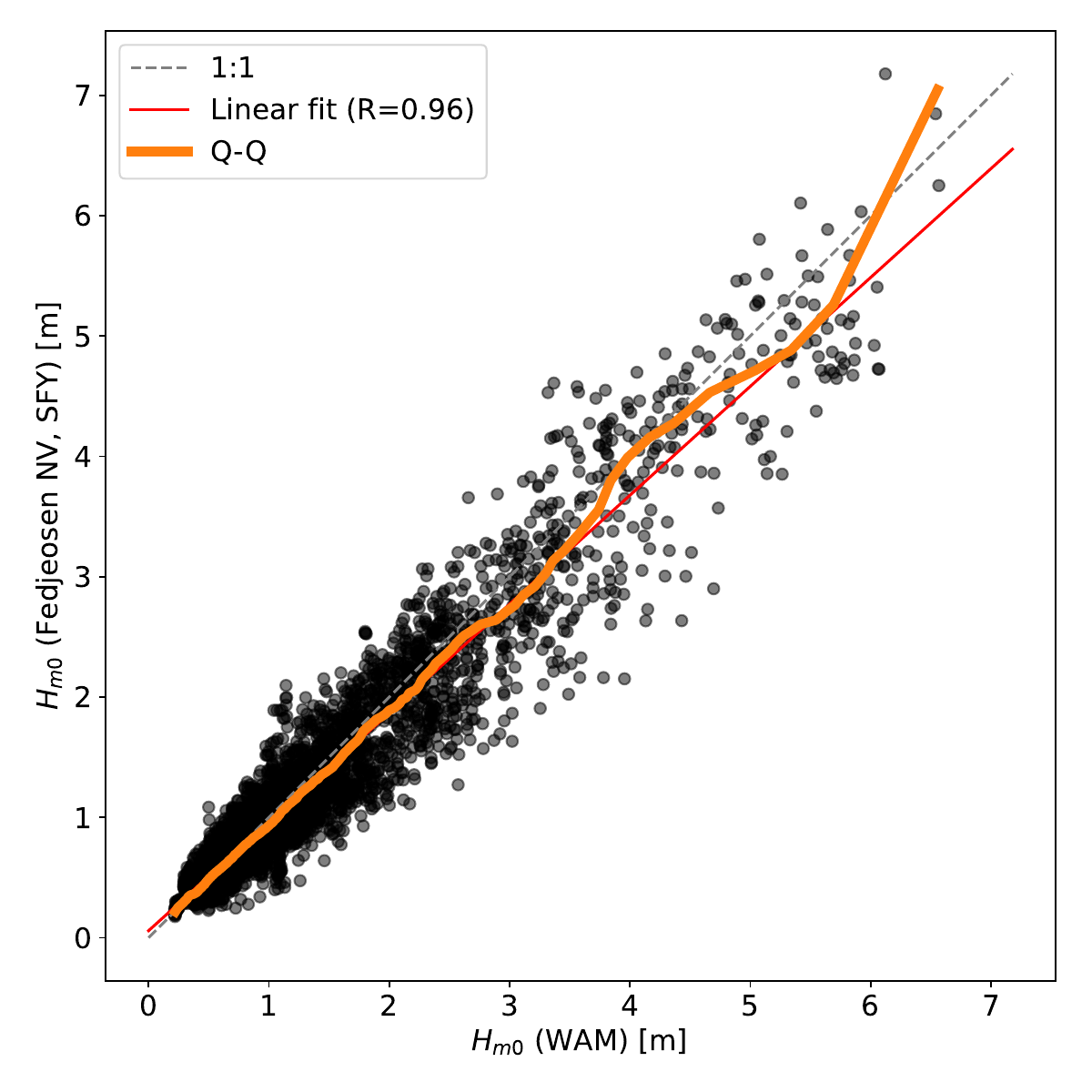}
  \caption{Observations from the buoy compared to the simulated $H_{m0}$ (WAM800 with current). Linear regression and quantile-quantile (Q-Q) lines are shown. The model predicts slightly higher significant wave height, but with an otherwise fairly linear relationship (Pearson correlation coefficient, $R = 0.96$). The model without current shows the same $R$-value.}
  \label{fig:fedjeosen-scatter}
\end{figure}


We use output from the operational coastal wave model WAM \citep{gunther1992WAMModelCycle,thewamdigroup1988WAMModelThird,komen1996DynamicsModellingOcean} as a reference for the wave conditions at the buoy site. The wave model WAM (cycle 4.7) is run on a rotated grid with 800~m grid spacing for five domains to cover the Norwegian coast. The model is run with ST4 physics \citep{ardhuin2010SemiempiricalDissipationSource} with a spectral resolution of 36 frequencies and 36 directions forced with surface winds from Arome \citep{muller2017AROMEMetCoOpNordicConvectiveScale}. The domains are nested into a WaveWatchIII\footnote{\url{https://github.com/NOAA-EMC/WW3.git}} setup on a large domain with 4 km grid spacing. A separate set of forecasts with WAM are run with input of ocean surface (0~m) currents from the ocean model ROMS\footnote{\url{https://github.com/myroms/roms.git}}. The archived forecasts from Arome, WaveWatchIII, ROMS and WAM can be accessed from an OPeNDAP server\footnote{\url{https://thredds.met.no/thredds/fou-hi/fou-hi.html}}. The first 12 hours of each WAM model run (00 and 12 UTC) at the nearest grid point is used to compile the time series of significant wave height.

The measured significant wave height is compared to the output of significant wave height from the WAM domain for the west coast of Norway, Vestlandet (Fig.~\ref{fig:fedjeosen_hm0} and~\ref{fig:fedjeosen-scatter}), with and without currents. The nearest grid point in the wave model is about $280$ m away from the buoy location. Figure~\ref{fig:fedjeosen-scatter} compares the observations against the simulated $H_{m0}$, and there is a good match with an $R$-value of $0.96$ for both models. The buoy measures slightly lower $H_{m0}$ values than the model predicts, except for a few instances of higher wave heights where the buoy measures higher than the model predicts. In general the relationship is close to linear.



Deploying the SFY at an exposed location like Fedjeosen provides the opportunity to compare its recordings with satellite observations of significant wave height to add an additional reference for the comparison. Measured $H_{m0}$ was compared to the near real time L3 satellite altimeter measurements from the Copernicus Marine Service catalogue product\footnote{\url{https://data.marine.copernicus.eu/product/WAVE_GLO_PHY_SWH_L3_NRT_014_001/description}} \citep{e.u.copernicusmarineserviceinformationcmemsmarinedatastoremds2023GlobalOceanSignificant}.
The satellite data are retrieved and collocated with the SFY using the open source tool wavy\footnote{\url{https://wavyopen.readthedocs.org}} following the collocation routine with temporal and spatial constraints as outlined by \citet{bohlinger2019NovelApproachComputing}. To acquire enough data, we set the temporal constraints to $\pm10$ minutes of the time stamp of the significant wave height, and a maximum spatial distance of $100$ km. This resulted in matches with 6 different satellites, namely Sentinel-3A/B, Cryosat-2, CFOSAT, SARAL-AltiKa, and Sentinel-6A Michael Freilich.
Satellite observations appear to follow closely the time series of $H_{m0}$ recorded by the SFY (Fig.\ref{fig:fedjeosen_hm0}).

\begin{figure*}[ht]
  \centering



  \includegraphics[width=\textwidth]{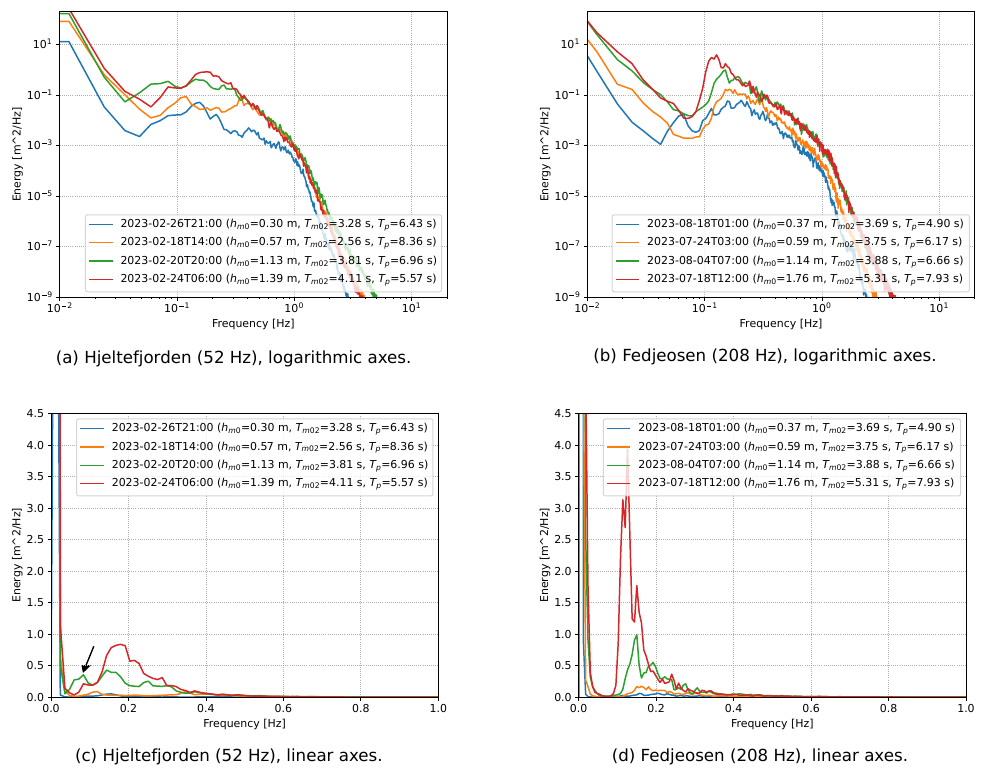}

  \caption{Elevation energy (variance density) spectra for comparable situations (similar $H_{m0}$ and $T_p$) of the same buoy (Fig.~\ref{fig:fedjeosen-mooring}) deployed at two different locations. The IMU (before feeding the AHRS-algorithm) was configured to sample at $52$ Hz in the first case (Hjeltefjorden), and at $208$ Hz in the second case (Fedjeosen). The arrow in the lower left panel shows low-frequency noise common to IMU buoys caused by insufficient sampling of impulsive events (see Sec.~\ref{sec:discussion}\ref{subsec:low-frequency-noise} for an extensive discussion).}
  \label{fig:open-water-spectra}
\end{figure*}

In Fig.~\ref{fig:open-water-spectra} spectra for 20-minute windows of the time series are compared to spectra for the same buoy deployed inside Hjeltefjorden a few months earlier. The deployment was very similar except that for the first deployment the IMU was sampling at $52$ Hz, while for the second (in Fedjeosen) the IMU was sampling at $208$ Hz and downsampled after being passed through the AHRS-algorithm. The spectra (Fig.~\ref{fig:open-water-spectra}) have been chosen for similar wave situation in terms of $H_{m0}$ and $T_p$. A notable difference is that for the buoy configured to sample at the lower frequency ($52$ Hz), more low-frequency noise is observed (arrow in Fig.~\ref{fig:open-water-spectra}c). This difference will be discussed more in Sec.~\ref{sec:discussion}\ref{subsec:low-frequency-noise}.

Between $1$ and $1.1$ Hz, after the $f^{-4}$ slope there is a pole (onset of increased negative logarithmic slope). This corresponds with the frequency of wavelength twice the horizontal diameter of the buoy ($65$ cm) which is $1.09$ Hz. It therefore seems that the dominant factor capping the upper measurable frequency is the size of the buoy.


\subsection*{Breaking waves measured by the moored buoy}
\label{subsec:fedjeosen-breaking}

\begin{figure*}[t]
  \centering



  \includegraphics[width=\textwidth]{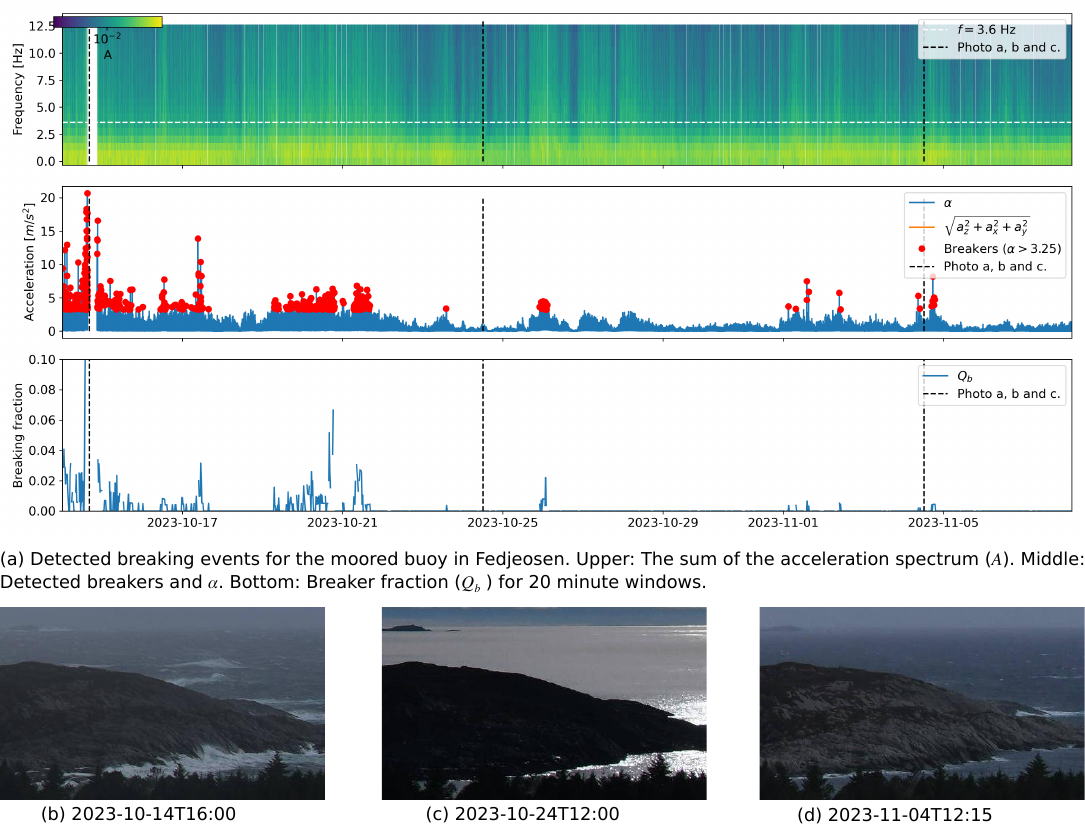}

  \caption{Breaking events and photos taken towards the buoy position (located to the left behind the hill) at times marked in (a). Photos by Norsk Luftambulanse, used with permission.}
  \label{fig:fedjeosen-break-photos}
\end{figure*}

Breaking waves hitting or hurling a buoy cause rapid acceleration impulses as can be clearly seen in the wave flume (Fig.~\ref{fig:breaker-lab}). A breaking wave strikes the buoy or flushes it sideways, causing an impulsive broad-banded signal that is not dependent on the geometric cut-off frequency. The measured strength of this impulse depends on buoy dimensions and wave steepness and how the breaking wave impacts the buoy, rather than the wave-length of the wave (before it breaks). \citet{seldal2023SFYFreedriftingWave} did a preliminary investigation into the signature of the breaking waves to try distinguish the different types of breakers in a way similar to the approach of \citet{brown2021AccelerationsWaveMeasurement}. The simplest method of quantifying the number of breakers would be to count maxima in the time series of horizontal acceleration, but this would detect many peaks for each breaking event. In order to make the method more robust against false positives we apply the method from \citet{brown2018KinematicsStatisticsBreaking}. Here we calculate the Short-Time Fourier-Transform (STFT) for tapered windows of $t_{\textrm{bw}} = 1.5$ s (overlap between windows of $t_{bw} / 2$) for all three components of acceleration ($a_z$, $a_x$, and $a_y$):
\begin{equation}
  A(f, t') = |\mathcal{F}\{a_z\}| + |\mathcal{F}\{a_x\}| + |\mathcal{F}\{a_y\}|
\end{equation}
and take the sum ($\alpha$) of the absolute magnitude along the frequency dimension and the sum ($A$) of components for frequencies above $f_{\mathrm{bc}} = 3.6$ Hz:
\begin{equation}
  \alpha(t) = \sum\limits_{f>3.6} A(f, t).
\end{equation}
Regular, linear waves, do not exist at significant energy levels at frequencies greater than $2$ Hz and we here assume that the energy found at these frequencies is caused by impulsive accelerations from breaking waves. The frequency was chosen so that it is above the consistent band of energy (upper panel in Fig.~\ref{fig:fedjeosen-break-photos}).

\begin{figure}[h]
  \centering
  \includegraphics[width=\hsize]{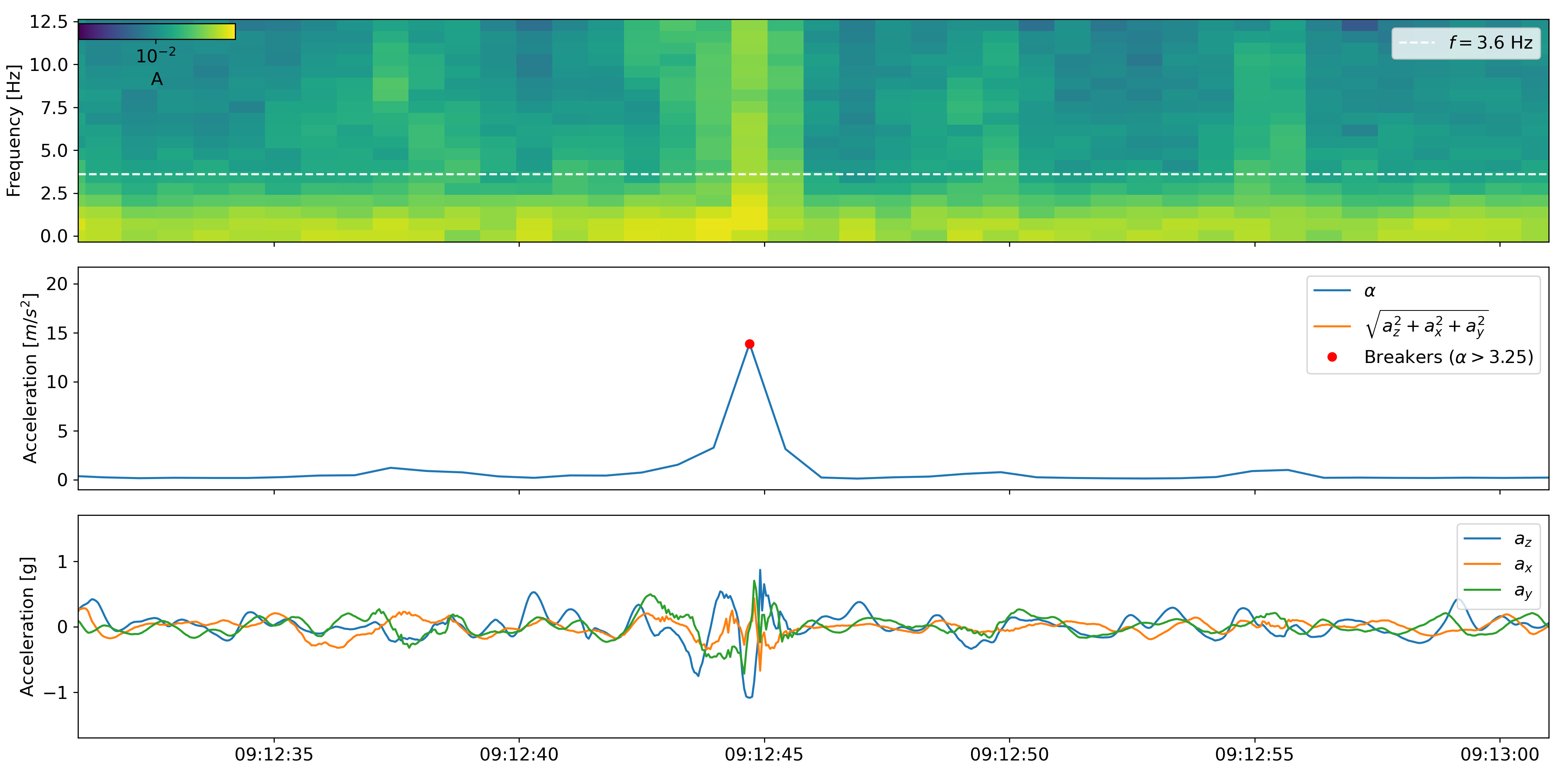}
  \caption{Detected breaking events for the moored buoy in Fedjeosen. \emph{Upper:} The sum of the acceleration spectrum ($A$). \emph{Middle:} Detected breakers and $\alpha$. \emph{Bottom:} Unfiltered acceleration components.}
  \label{fig:fedjeosen-break-single-detail}
\end{figure}

Breakers are identified by counting peaks of $\alpha$ above a threshold value. The parameters chosen for window length ($t_{\mathrm{bw}}$), $f_{\mathrm{bc}}$ and the threshold for $\alpha$ are dependent on the buoys' response to breakers, and signal processing \citep{brown2018KinematicsStatisticsBreaking}. As in \citet{brown2018KinematicsStatisticsBreaking} we use a threshold $\alpha = 3.25\,\mathrm{m}\,\mathrm{s}^{-2}$, and rather increase the cut-off frequency to $3.6$ Hz with the argument that only counting high-frequency impulses is a more robust way of detecting the breakers. Along with the plot of $\alpha$ and the detected breakers in the middle panel in Fig.~\ref{fig:fedjeosen-break-photos}, we also plot the norm of the acceleration components $\sqrt{a_z^2 + a_x^2 + a_y^2}$, filtered above $2$ Hz. It gives very similar results to $\alpha$, with the main difference caused by the smoothing due to the windowing in $\alpha$ (middle panel in Fig.~\ref{fig:fedjeosen-break-single-detail}).


In Fig.~\ref{fig:fedjeosen-break-single-detail} broadbanded maxima in acceleration (visible as vertical stripes in the acceleration spectrum) are used to detect breakers. In the lower panel the three acceleration components are shown (unfiltered). Similar impulses to the lab experiments are apparent (Fig.~\ref{fig:breaker-lab}).

A way to quantify breaking waves is to estimate the breaking fraction, the fraction of waves passing a point that are breaking (or approximately breaking probability) \citep{schwendeman2014WaveBreakingDissipation,brown2018KinematicsStatisticsBreaking},
\begin{equation}
  Q_b = N_b T / \tau.
\end{equation}
Here, $N_b$ is the number of breakers detected during a measurement interval, $\tau = 20$ min, and $T$ is the wave period, here approximated with $T_{m01}$. Figure~\ref{fig:fedjeosen-break-photos}a shows $Q_b$ (lower panel) for a subset of the data. We have visually inspected images taken every 15 minutes with an air traffic camera on the Fedje Island pointing in the direction of the wave buoy. In Fig.~\ref{fig:fedjeosen-break-photos} photos from three instants are shown. These times are marked in Fig.~\ref{fig:fedjeosen-break-photos}a with vertical dashed lines. Qualitatively, the detection of breakers agrees well with the visual impression of breaking waves in the photographs. The wave steepness ($H_{m0} / (g \cdot \frac{T_{m01}^2}{2\pi})$) for the measurement period varies very little (but agrees well with models).


\section{Arrays of wave buoys in the surf}
\label{sec:surf-experiments}

The buoy was tested in the surf in two separate locations for two types of experiments. First to record the waves and breaking events simultaneously on three buoys (Jæren, Norway), and demonstrate that it is possible to study the phase difference of the events between the buoys in the array. Secondly, to try to measure the trajectory of a single buoy as it is caught in a breaking wave (Green Bowl, Bali, Indonesia).

\subsection{Array of buoys in the surf}
\begin{figure}[h]
  \centering

  \includegraphics[width=\hsize]{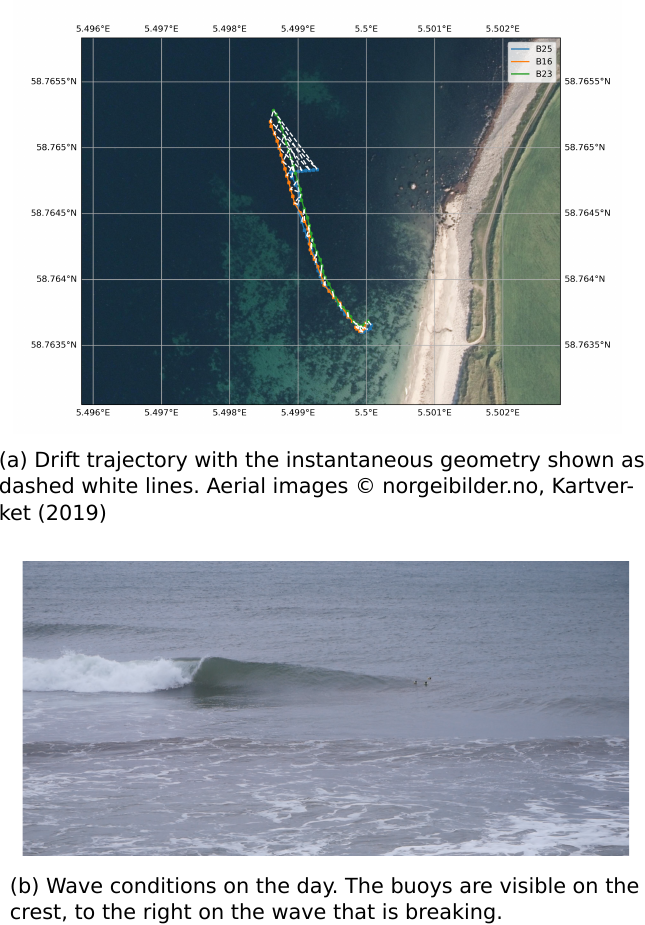}
  \caption{Free-drifting array of buoys in the surf outside Jæren.}
  \label{fig:map-svini}
\end{figure}

The first three buoys were deployed as a free-drifting array on the beach outside Jæren on the west coast of Norway on the 8th of January 2023. The cross-shore wind and rip currents carried the buoys outwards and they needed to be picked up before drifting too far out at sea. Figure~\ref{fig:map-svini} shows the drift trajectory of one of the deployments and the conditions of the day. As the buoys move independently of each other, their relative positions are continuously changing (white dashed triangles).

\begin{figure}[h]
  \centering
  \includegraphics[width=\hsize]{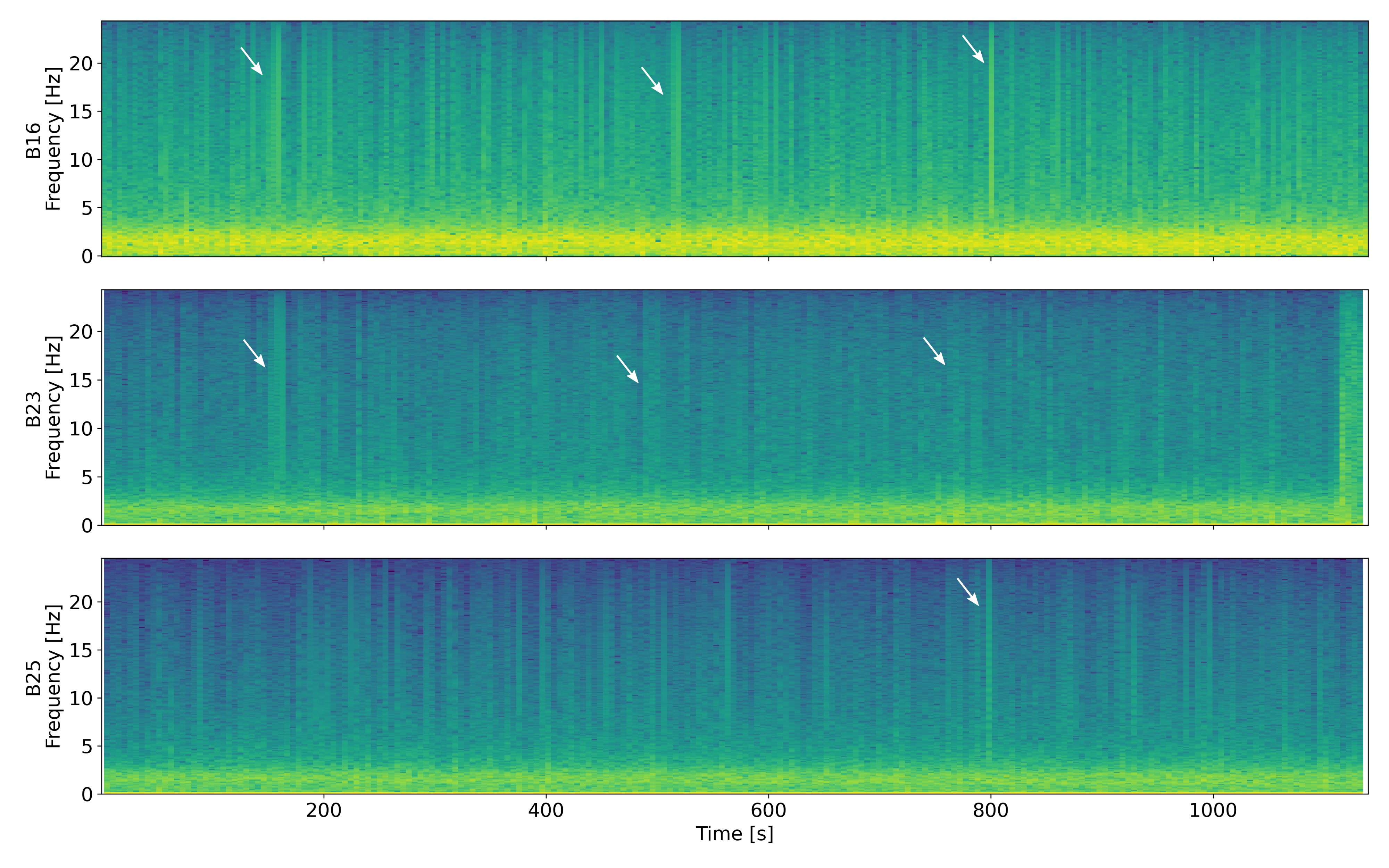}
  \caption{Vertical acceleration spectrograms for three buoys, captured in the surf at Jæren. Vertical lines of high intensity signals (marked with arrows) spanning the full frequency spectrum is visible at each buoy, with some delay in phase.}
  \label{fig:svini-spec}
\end{figure}

Figure~\ref{fig:svini-spec} shows the acceleration density spectrogram for the vertical component of the three buoys for drift shown in Fig.~\ref{fig:map-svini}. Vertical lines of high energy impulses assumed to be caused by breaking events are visible, and are correlated in time between the buoys. Since the buoys in this experiment were configured with very sensitive accelerometer range ($\pm 2g$) and gyroscope range ($\pm 125$ dps), the sensor saturates and the trajectories are difficult to recover.

\subsection{Trajectory in a breaking wave}

\begin{figure}[h]
  \centering
  \includegraphics[width=\hsize]{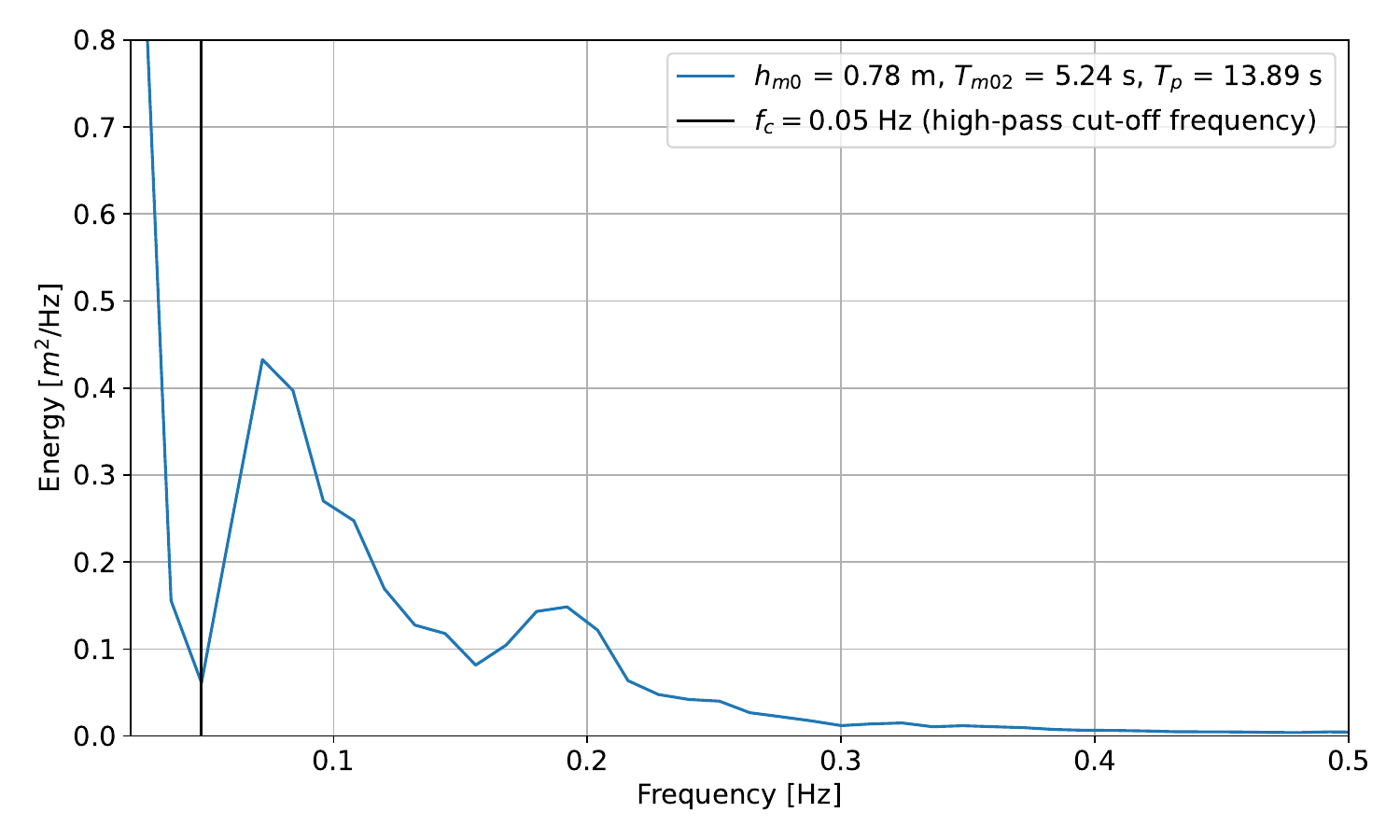}
  \caption{Spectrum captured outside the surf at Green Bowl, Bali. The vertical line shows the high-pass cut-off frequency ($f_c$) to avoid the low-frequency noise common to IMU buoys (see Sec.~\ref{sec:discussion}\ref{subsec:low-frequency-noise} for an extensive discussion).}
  \label{fig:spectrum-green-bowl}
\end{figure}

In a similar experimental setup, a single SFY buoy was deployed in the surf at Green Bowl, Bali, Indonesia, in November 2023. In this deployment the acceleration range is set to $\pm 16g$ and the gyroscope range is set to $\pm 1000$ dps, and the AHRS filter is run at $208$ Hz. First the spectrum (Fig.~\ref{fig:spectrum-green-bowl}) was measured for 20 minutes outside the breaking waves as a reference. The swell was about 0.9 m while the surfing wave height was about 0.3--0.5 m. The spectrum does not consist of a single peak, so irregular sets of waves breaking onto the shore is expected.

The buoy was then deployed so that it would drift towards the shore through the breaking waves. The horizontal components are re-projected along the direction of maximum variance as in Sec.~\ref{sec:lab-experiment}\ref{subsec:lab-traj} and integrated to the horizontal and vertical displacement. A cut-off frequency of $0.04$ Hz is used when filtering before integrating.

In Fig.~\ref{fig:green-bowl-break-photos} a wave approaches the buoy (painted red). The buoy is drawn offshore towards the surf wave \emph{(a)}, and almost escapes the breaking crest \emph{(b)}, but eventually the buoy is caught in the overturning wave \emph{(c)} and thrown towards the surface of the water (exceeding $10 g$ upwards acceleration on impact) and flushed out. A second orbit begins as the next (non-breaking) wave travels towards the shore.

\begin{figure}[h]
  \centering
  \includegraphics[width=\hsize]{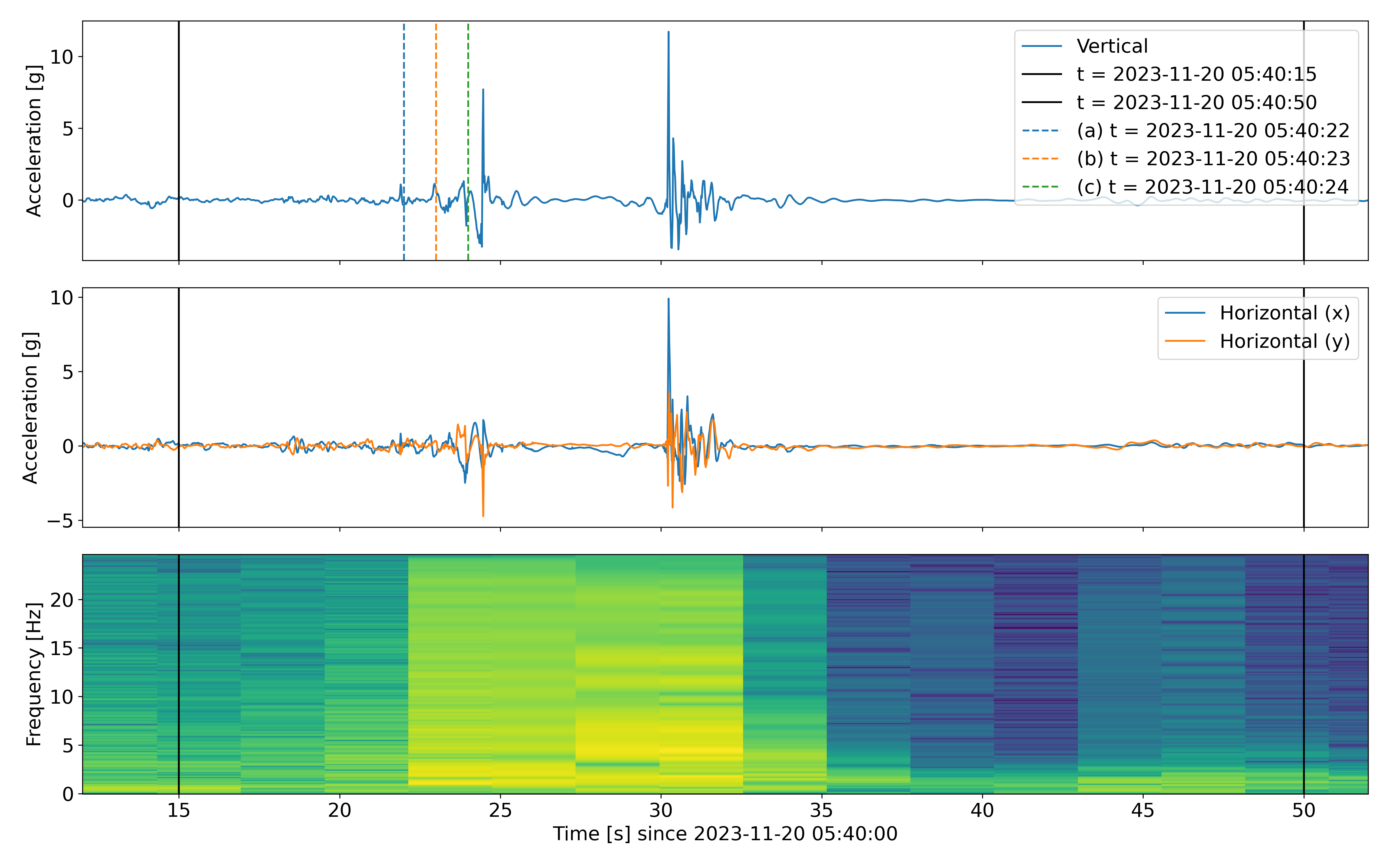}
  \caption{Acceleration and spectrogram at Green Bowl, Bali of a buoy in the breaking waves of the surf.}
  \label{fig:trace-green-bowl}
\end{figure}

\begin{figure*}[t]
  \centering


  \includegraphics[width=\textwidth]{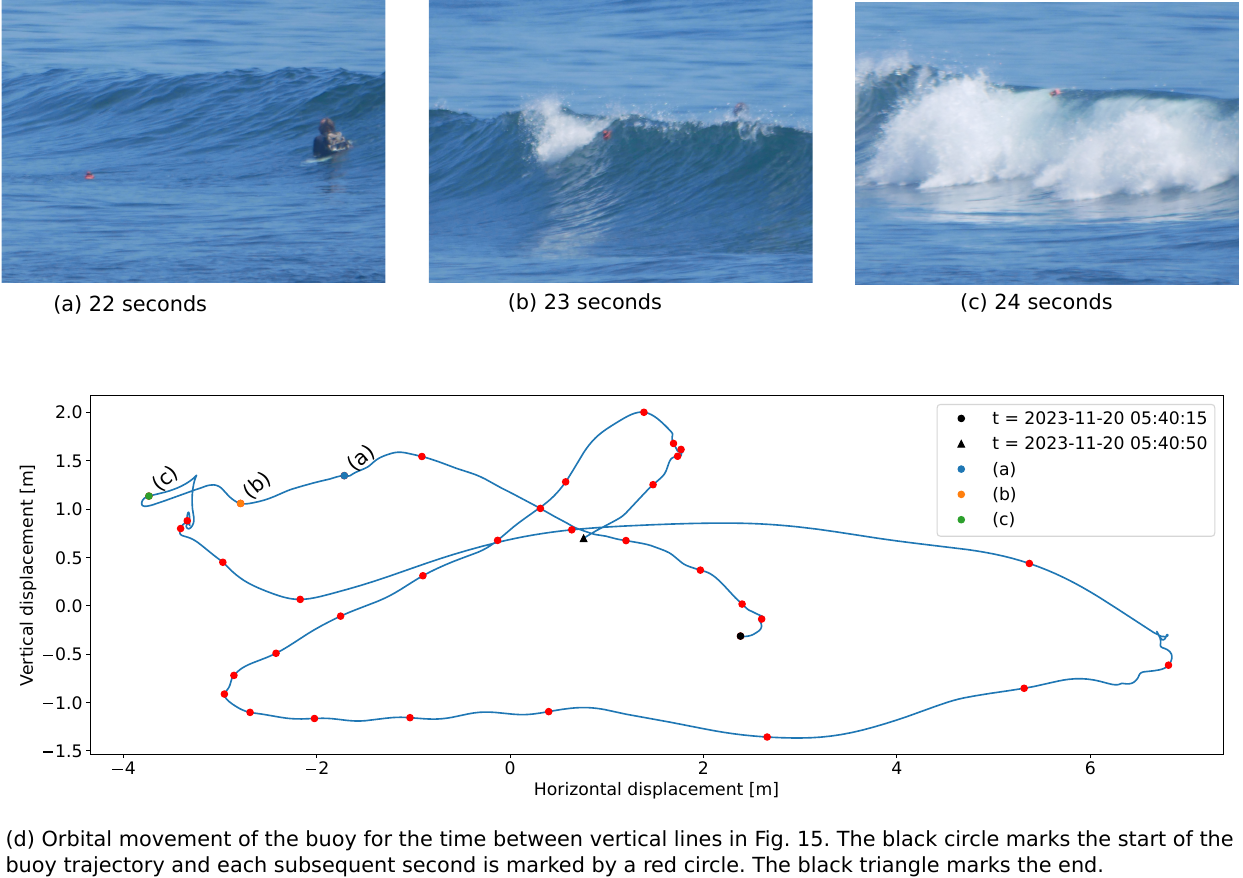}
  \caption{Orbital movement of buoy and photos captured at the same breaking wave recorded by the buoy in Fig.~\ref{fig:trace-green-bowl}.}
  \label{fig:green-bowl-break-photos}
\end{figure*}


The recorded displacement amplitudes match well with expected values from the spectrum and visual observations. Usually, measurements from breaking waves in the surf are not suitable to be integrated to displacement without heavy filtering. In Fig.~\ref{fig:green-bowl-break-photos} we can match the photographs with the integrated trace (Fig.~\ref{fig:trace-green-bowl}), and see that the recorded trajectory (Fig.~\ref{fig:green-bowl-break-photos}d) is sensible even with a very low cut-off frequency. We are at the limit of what trajectories are possible to reconstruct, therefore the trajectory may be tilted due to insufficient speed of the AHRS-algorithm. The time of the images may also be slightly shifted, as synchronization is insufficiently accurate. This is discussed further in Sec.~\ref{sec:discussion}\ref{subsec:disc-breaking}.

\section{Discussion}
\label{sec:discussion}

\subsection{Low-frequency noise and breaking waves}
\label{subsec:low-frequency-noise}


All IMU-based wave buoys will struggle to measure low-frequency signals. The sensor measures acceleration, and the double integration to obtain vertical displacement (Eqs.~\eqref{eq:int-v} and \eqref{eq:int-eta}) introduces drift which must be removed using detrending. Additionally, a core assumption of AHRS-algorithms is that they measure a high-frequency signal, and very long period waves may not fall into this domain and appear to shift the Earth's gravity vector from the buoys point of view.

The lower frequency signal and noise of the acceleration are amplified compared to the higher frequencies through the integration, Eqs.~\eqref{eq:elevation-freq} and \eqref{eq:elevation-energy-freq}. Displacement measurements from IMUs are therefore inherently sensitive to low-frequency noise. The low-frequency noise can be seen in Figs.~\ref{fig:lab-welch-spectrums} and~\ref{fig:open-water-spectra}, and is similar to spectra measured by other IMU-based buoys \citep{nose2023ComparisonOperationalWaveice,chuang2013ContinuousWaveletTransform,arraigada2006CalculationDisplacementsMeasured,collins2014RecordingSeaSurface,brown2018KinematicsStatisticsBreaking, verasguimaraes2018SurfaceKinematicsBuoy, rabault2020OpenSourceVersatile, rainville2022MeasurementsNearshoreWaves}. The longest wave periods that can be measured robustly using IMUs are about $20$ seconds ($0.05$ Hz), depending on the conditions \citep{brown2018KinematicsStatisticsBreaking,vanessen2018WaveBuoyPerformance,datawellbv2023DatawellWaveriderManual}. Note that up to $100$ s periods can be observed by using e.g. GPS to measure wave orbital velocities, although this is also not trivial \citep{thomson2018MeasurementsDirectionalWave,collins2024PerformanceMooredGPS}. Note that when comparing spectra the length of the signal should be similar in order to expect similar signal to noise ratio.

Consequently, the acceleration must be high-pass filtered with a chosen cut-off frequency. Choosing this cut-off frequency automatically is not trivial, since the assumed low-frequency noise often extends higher than $0.05$ Hz and can be difficult to distinguish from the signal. In this analysis we use a relatively simple algorithm where we attempt to find a minimum in the smoothed spectrum above $0.04\textrm{--}0.05$ Hz \citep{nose2018PredictabilityStormWavea,rabault2022OpenMetBuoyv2021EasytoBuildAffordable} which is used as the cut-off frequency (see Sec.~\ref{sec:lab-experiment}\ref{subsec:lab-significant-wave-height}).

In particular, in experiments in the surf-zone we notice that there is often no significant energy minimum between the main peak ($f_p$) and the low-frequency noise. The same has been observed for sporadic 20-minute periods in the SFY and OpenMetBuoy when deployed in the open ocean (also observed in other buoys \citep{nose2018PredictabilityStormWavea}). These spectra tend to give exaggerated estimates for $H_{m0}$ or invalid estimates for $T_p$ \citep{alari2022LainePoissLightweightIceResistant}.

Most other assessments of the performance and behavior of IMU based buoys observe and identify the low-frequency noise (see references above), but its origin is not well explained. Some of it is attributed to the integrated noise of the IMU. However, that does not explain why there is much higher low-frequency noise in some situations than others, or that the noise is well above the expected noise-floor (when extrapolating the noise curve from high to low-frequencies). \citeauthor{alari2022LainePoissLightweightIceResistant} concluded that the low-frequency energy is misplaced from other frequencies, and that it should not be filtered completely away for calculation of spectral moments as is the usual way to deal with this noise. \citet{brown2018KinematicsStatisticsBreaking} also attribute some of the low-frequency noise as aliasing from higher frequencies. For the MOTUS buoy \citep{dorgeville2018DataValidationMOTUS}, a mechanical filter was used to reduce measured vibrations (from wind, waves, and currents) above the Nyquist-frequency that fold back into lower frequencies. However, for the SFY both the gyroscope and accelerometer have built-in low-pass filters that would \emph{avoid sampling alias}.

Another concern is that rotation causes centrifugal acceleration. \citet{feddersen2023WavedrifterLowcostIMUbased} found that a buoy caught in a overturning breaking wave, with the IMU located fairly close to the center, only undergoes small centrifugal accelerations. 
They find that angular velocity of a tumbling buoy very seldom exceed $600$ dps during breaking events, and never more than $1000$ dps. The SFY buoy has sampled at ranges of $\pm 125$ and $\pm 500$ dps in the open-water, and $\pm 1000$ dps in the surf zone (a lower range gives a better resolution, unless it saturates).

We observe that the low-frequency noise occurs more frequently in the surf, and only sporadically in the open ocean. As wave breaking is more violent and happens much more often in the surf zone, it seems likely that high-frequency acceleration impulses from breaking waves are the cause of the low-frequency
noise in the measured elevation. Specifically, rapid acceleration and rotation caused by a breaking wave insufficiently sampled into the fusion algorithm would cause the buoy to lose orientation and no longer correctly extract the vertical acceleration. To test this hypothesis we configured buoys with different IMU sample rates ($208$ Hz and $52$ Hz), but equal output rate (filtered using a FIR-filter and decimated). In the wave flume (Sec.~\ref{sec:lab-experiment}), both the low-frequency and the high-frequency buoys performed well on harmonic waves (Fig.~\ref{fig:lab-traj-uz}). However, when the buoys are subjected to breaking waves in the wave flume (Sec.~\ref{sec:lab-experiment}\ref{subsec:lab-breaking-waves}), the buoy with the higher IMU-sampling (B25) performs far better than the other (B23), with low-frequency oscillations clearly visible.

In fact, a similar phenomenon is observed in GPS-based buoys where discontinuities in the time-series manifest as low-frequency noise \citep{bjorkqvist2016RemovingLowfrequencyArtefacts,collins2024PerformanceMooredGPS}.

A buoy that encounters breaking waves only sporadically yields a spectrum  contaminated by these impulses. If this is the case, higher sample rate should better resolve the rotation and acceleration and thus result in less low-frequency noise. The spectra in Fig.~\ref{fig:open-water-spectra} suggest that the buoy with $208$ Hz sampling and AHRS frequency performs better on low-frequencies than the one with $52$ Hz sample rate. Detecting and removing segments with breaking impulses from the time series should also improve the spectral estimates.

Additionally, a buoy in breaking waves requires much higher sample rates and ranges in measuring rotation and acceleration. Given the large accelerations measured in breaking waves (several $g$ in the open water, and more than $10g$ in the surf in our case (Sec.~\ref{sec:surf-experiments}), and $8\textrm{--}14$  reported by \citet{sinclair2014FlowRiderLagrangianFloat}, \citet{brown2018KinematicsStatisticsBreaking}, and \citet{feddersen2023WavedrifterLowcostIMUbased}), a much higher sample rate is probably required to have a chance at measuring these signals with any success.


\subsection{Capturing and quantifying breaking waves}
\label{subsec:disc-breaking}

A breaking wave is a highly impulsive event. Impulses have infinite bandwidth, and it is impossible to fully capture them with a band-limited method. Our experiments in the wave flume (Sec.~\ref{sec:lab-experiment}\ref{subsec:lab-breaking-waves}), the open water (Sec.~\ref{sec:fedjeosen}) and in the surf (Sec.~\ref{sec:surf-experiments}) show that the horizontal and vertical acceleration impulse of a breaking wave far exceed those of regular waves. The acceleration experienced by the buoy during a breaking event exceed $10 g$, consistent with previous work \citep{sinclair2014FlowRiderLagrangianFloat,brown2021AccelerationsWaveMeasurement,feddersen2023WavedrifterLowcostIMUbased}. An important observation is that the high-acceleration impulses are very short and they will not be adequately captured unless sampled frequently enough.

In the open water (Sec.~\ref{sec:fedjeosen}) the breaking fraction is quantified using methods developed by \citet{brown2018KinematicsStatisticsBreaking}. The fraction qualitatively agrees with photos from the location. However, a more controlled experiment with camera on the buoy, or the buoy located in the field of vision of stereo-video cameras, would provide more reliable context. This would allow better evaluation of how robust this method is and the influence of the parameters that go into it.

\citet{feddersen2023WavedrifterLowcostIMUbased} captures the trajectory of a breaking wave generated at the \emph{Surf Ranch} remarkably well. In this work we attempt to capture the trajectory in the waves breaking onto a beach (Sec.~\ref{sec:surf-experiments}). The waves breaking on the beach do not consist of a single reproducible, relatively clean, wave. Yet, we do capture the trajectory well compared to the spectrum outside the surf and the photographs. The signal was filtered with a cut-off frequency as low as $0.04$ Hz, without resulting in exaggerated displacements. To our knowledge a breaking wave in the ocean has not been captured with a buoy in this way before.

The reason that the buoy can measure this way is presumably due to a combination of using a small wave buoy, high frequency sampling, and a wide measurement range on the IMU (accelerometer and gyroscope).

\subsection{Cellular network and power source}

The cellular network has performed remarkably well even when the antennas have only been a few cm above the surface. A better container with a better aligned and placed antenna is likely to perform better. In the moored buoy, where the antenna can be placed higher ($30 \textrm{--} 40$ cm), the modem connects with a cellular tower $80$ km away (Fig.~\ref{fig:fedjeosen-tower-range}). Only during the largest waves ($H_{m0} > 7$ m) do we see intermittent failures in transmitting data. This will cause loss of data when the buoy does not have a large enough internal buffer. A larger buoy might be more stable in this situation and perform better. It is likely that also satellite-based telemetry will struggle to transmit data in such sea states.

\section{Conclusions}
\label{sec:conclusions}

The SFY, a small, light-weight, drifting or moored wave buoy design with the capability to detect breaking events has been described. The buoy is designed to work in coastal environments and communicates over the cellular network as well as storing the data on SD-card. Its design, code, and hardware is open source and available publicly\footnote{\url{https://github.com/gauteh/sfy}}. The buoy has been proven to work reliably up to $80$ km from the nearest cellular tower, and in its small form measures frequencies from about $0.04$ Hz to about $2.2$ Hz (or $1.1$ Hz for the moored buoy, without requiring correction). The main advantages and improvements of the SFY buoy are:
\begin{itemize}
  \item \emph{light-weight} ($\approx 0.5$ kg) and \emph{low-cost} ($\approx 200$ USD): allowing many buoys to be deployed, easy deployment from small vessels or even by swimmers or surfers.
  \item \emph{high-bandwidth and efficient processing}: with a greater bandwidth for data transmission, the full time series of AHRS-processed \emph{high-frequency} accelerations ($52$ Hz) is transmitted home, significantly more than for existing buoys. This allows a much more detailed on-land analysis of time-series, \emph{individual phases and breaking waves}, compared to transmitting only the wave spectra.
  \item \emph{open source}: anyone can build and modify the buoy to adapt it to their experiment, or use it to develop the next generation of buoys.
\end{itemize}

Through wave flume experiments, deployments in the open-water, and in the surf, the buoy is shown to measure significant wave height and other spectral properties accurately. It is shown to measure the surface elevation in such detail that the individual wave phases can be compared between adjacent buoys. These properties enable the buoy to be used in array configurations where the spatial details of the wave field should be studied, e.g., within a stereo-video footprint \citep{malilamika2022InvestigationDynamicalStatistical} or in areas with strong wave-current interactions.

The most challenging measurements were performed in the surf zone. The irregular waves breaking on the beach were not produced by a singular wave component or soliton. Yet, by measuring at very high frequency ($208$ Hz) and with a large dynamic range ($\pm 16g$ and $\pm 1000$ dps), we were able for \emph{the first time to reconstruct the trajectories} of the buoy through breaking and non-breaking waves on the beach to a high degree. Previously this has only been accomplished in artificial waves or through imaging methods. Necessary to this is that we were able to use a low cut-off frequency of $0.04$ Hz here, normally reserved for open-waters. The lack of exaggerated elevation amplitudes after integrating to displacement gives us confidence that the buoy measures correctly.

In Table~\ref{tab:buoy-config-scenario} we recommend parameter ranges depending on the deployment scenario.

\begin{table}[ht]
  \centering
  \begin{tabular*}{\hsize}{lcccc}
    \hline\hline
    Scenario   & Frequency & Acceleration & Gyroscope \\
    \hline
    Sea-ice    & $100$ Hz       & $\pm 2g$           & $\pm 125$ dps  \\
    Open-water & $208$ Hz       & $\pm 4g$           & $\pm 500$ dps  \\
    Surf       & $>208$ Hz      & $>\pm 16g$          & $\pm 1000$ dps \\
    \hline\hline
  \end{tabular*}

  \caption{Recommended sample rate and IMU range for different scenarios (see also \cite{sinclair2014FlowRiderLagrangianFloat,rabault2020OpenSourceVersatile,feddersen2023WavedrifterLowcostIMUbased}).}
  \label{tab:buoy-config-scenario}
\end{table}

An important conclusion which we surmise will be valid for all IMU-based buoys is that low-frequency noise can to a large degree be explained by impulsive events, like breakers, that are not adequately sampled. This can partly be attributed to aliasing of the large, but very short, acceleration peaks. However, the inadequate sampling is causing the AHRS algorithm to loose track of orientation and therefore recording horizontal movement in the vertical direction and vice versa. We find that by sampling the IMU and running the AHRS-algorithm at a higher frequency before downsampling its output (if bandwidth is limited), much better noise characteristics are achieved. By this logic, better noise characteristics or estimates based on the spectrum (like $H_{m0}$) may improve if impulsive events can be detected and excluded from the spectrum.

The advent of small, disposable drifting wave buoys with telemetry opens new opportunities for dense sampling of the surf zone and regions with strong spatial variability, such as tidal currents in open-ocean conditions (see, e.g., the studies of the wave modulation by the Moskenes tidal current in Northern Norway by \citealt{saetra2021IntenseInteractionsOcean,halsne2022ResolvingRegionsKnown,halsne2023WaveModulationStrong,halsne2024WaveModulationStrong}). Deploying large, densely spaced, arrays of buoys that sample at rates high enough to identify breakers and the accelerations in breaking events should allow for experiments that are currently not possible, or prohibitively impractical and expensive to do.

\acknowledgments

Jim Thomson, Alex de Klerk, E.J. Rainville and Falk Feddersen are thanked for
discussions and generously sharing their knowledge and experience. The
following are thanked for their help in deploying buoys in the surf and
technical assistance: Silje Røe, Susanne Moen Olsen, Judith Thu Ølberg, Daniel
Benedet, Max Deffenti, O.J. Aarnes. Jan Bjarte Valsvik and the students of
Jæren FHS are thanked for their help in the early experiments. Ragnhild and
Ingrid helped test the buoyancy, durability, and the first concepts of the
buoy.

This research and data collection has been supported by the Research Council of
Norway grants \emph{ENTIRE} (no. 324227), \emph{SFI BLUES} (no. 309281)  and
\emph{Havvarsel} (no. 310515), as well as by the Norwegian Coastal
Administration. P.B. gratefully acknowledges funding through the project
\emph{COSY – Coastal Operational and Sustainable use of satellite altimetrY},
funded by the Norwegian Space Center under grant 74CO2222.

%
%
\datastatement

The embedded code, bill of materials, tutorials for building the buoys, hardware schematics and circuit board files are available online at: \url{https://github.com/gauteh/sfy}. The processing code and server side code is available at the same place.

The data from Sec.~\ref{sec:fedjeosen} is available in CF-compliant NetCDF at: \url{https://thredds.met.no/thredds/catalog/obs/fouombuoys/fedjeosen_nv/catalog.html}. The data from Sec.~\ref{sec:lab-experiment} and~\ref{sec:surf-experiments} is available at \url{https://github.com/gauteh/sfy-2024-data}.


\appendix[A]
\appendixtitle{Power usage and cellular network range}

\section{Power usage}
\label{app:power}
\begin{figure}[H]
  \centering
  \includegraphics[width=\hsize]{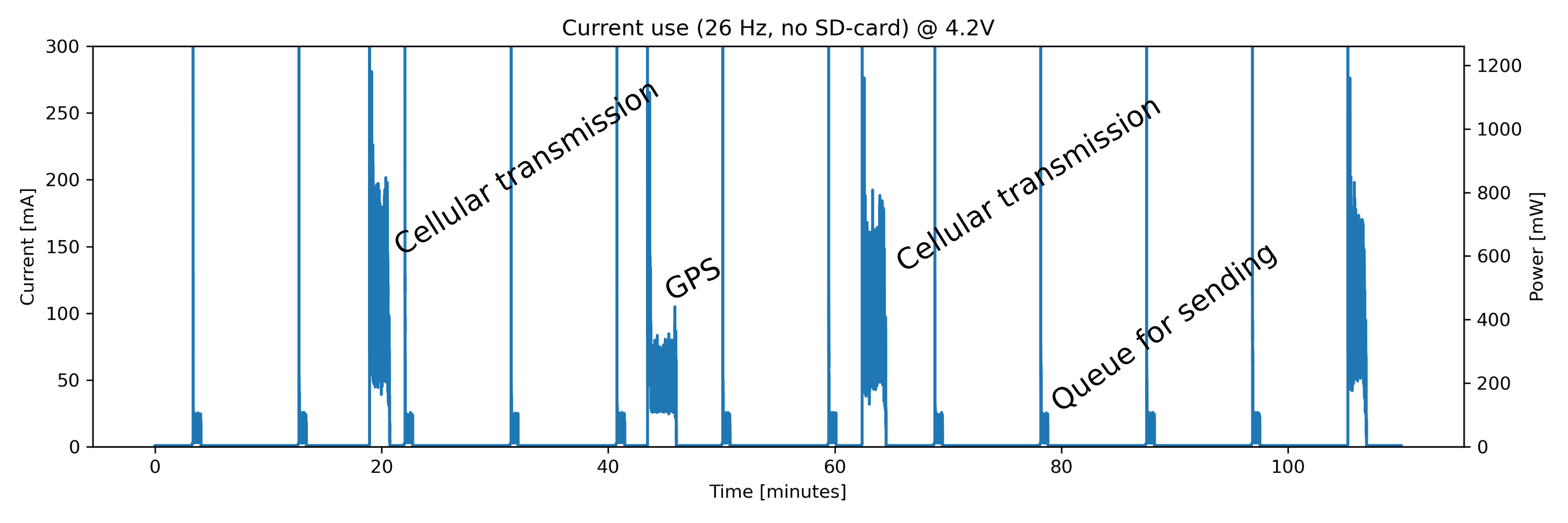}
    \caption{Current measured using a \emph{CurrentRanger}. The periods with the lowest current happen when the buoy is logging the IMU continuously, while not doing anything else. Higher current is drawn when the GPS or cellular communication is active. Spikes seen at the start of new communication or GPS periods are measurement errors due to input-range change from $\mu$A to mA in the \emph{CurrentRanger}.
    }
  \label{fig:power-current-ranger}
\end{figure}

The majority of the power is drawn by the cellular modem during connection and transmission of data. When the modem is connecting, current spikes up to $2$ A can occur for a few milliseconds at a time. Super-capacitors limit this to about $0.3$ A, and thus enable batteries with higher internal resistance to be used. Figure~\ref{fig:power-current-ranger} shows the measured current (at $4.2$ V) during cellular connection, transmission, and lower-current logging for a buoy configured to log at $26$ Hz and with storage on the secure digital (SD) card temporarily disabled (as in Section~\ref{sec:fedjeosen}). The average power consumption for this configuration is about $32$ mW with an average current going as low as $0.6$ mA when logging to memory only. Note that spikes at the beginning of each section are artefacts caused by measurement range change in measuring the current.

Considerable effort has gone into making the buoy work well with alkaline batteries rather than lithium-based batteries. Lithium batteries can deliver much greater power and capacity, but mixing lithium batteries with water is not safe \citep{hasvold2007SafetyAspectsLarge}. Since this buoy is designed to work close to the coast and to be drifting freely there will always be an appreciable chance that the buoy is found and picked up by the public. This puts stricter requirements on power-use and maximum current draw.

We found the buoys lasted  about $12$ days with continuous logging and transmission during field experiments with a logging frequency of $52$ Hz, storage to an SD card and frequent GPS fixes. This autonomy was obtained with 6 alkaline C-cells with individual capacity of 8 Ah @ 1.5V \citep{duracell2019DuracellProCellIntense}, or 16 Ah @ 4.5 V with $2 \times 3$ batteries. That amounts to an average current use of about $\mathrm{16 Ah / (12 \times 24) h = 0.055 A = 55 mA}$. However, when the buoy is configured to take positions every hour, without an SD-card, and with a lower output frequency, the current drain is reduced down to about $7.8$ mA (or $32.8$ mW).

A moored version of the buoy (as described in Section~\ref{sec:fedjeosen}) with $3\times 21$ alkaline D-cells (21 parallel $4.5$ V) shows an initial voltage drop of $2.47$ mV on average each day for the first $60$ days. The voltage drop of alkaline cells is steeper in the beginning and the end the battery capacity. A conservative estimate of the battery capacity with a working voltage between $4.621$ and $3.650$ V, yields a battery time of about $390$ days.

\section{Cellular network range}
\label{app:cellular}

\begin{figure}[H]
  \centering
  \includegraphics[height=4cm]{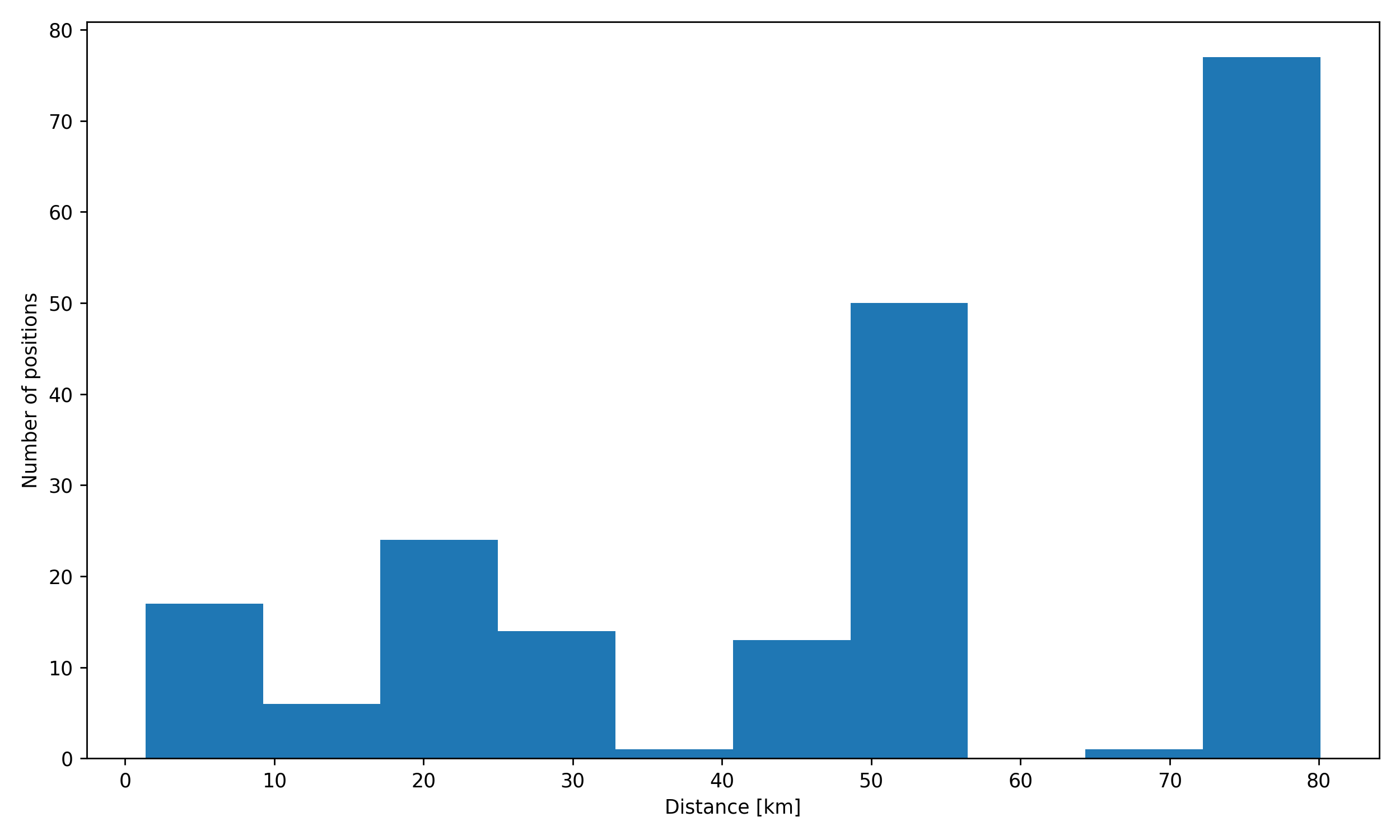}
  \caption{The distance to the cellular tower from the buoy.}
  \label{fig:fedjeosen-tower-range}
\end{figure}

Figure~\ref{fig:fedjeosen-tower-range} shows a histogram of the distance to cellular tower used by the moored version of the buoy. Most of the time it uses a tower about $80$ km away.

\appendix[B]
\appendixtitle{Data transmission}
\section{IMU sampling and data collection}
\label{app:processing}

Due to chip-to-chip differences every IMU samples at a slightly different frequency (typically within $\pm 10\%$). The actual sample rate can be retrieved from the IMU registers \citep{stmicroelectronics2021AN5398ISM330DHCXAlwayson} (section 6.4). However, since the data is accurately timestamped using GPS, we estimate the true frequency accurately by the median time difference between every 1024-sample batch. This is done routinely to ensure correct sampling frequency.

The MCU configures the IMU to record acceleration and rotation into its first-in, first-out (FIFO) queue continuously, and then drains and processes the samples at sufficiently frequent intervals for the FIFO to not fill up. The MCU sleeps in between, saving power.

The data are collected in batches of 1024 samples and transmitted in messages tagged with position and timestamps. These data are relayed to a server that stores the messages in a database. A set of processing scripts read all messages in a given range and concatenate the data and construct a continuous time coordinate. The data are stored in CF-compliant NetCDF or zarr files with all metadata, and should be fully self contained and ready for further analysis.

\newpage
\appendix[C]
\appendixtitle{Schematic of mooring}
\section{Fedjeosen mooring}
\label{app:fedjeosen-mooring}

\begin{figure*}[ht]
  \centering
  \includegraphics[width=0.75\textwidth]{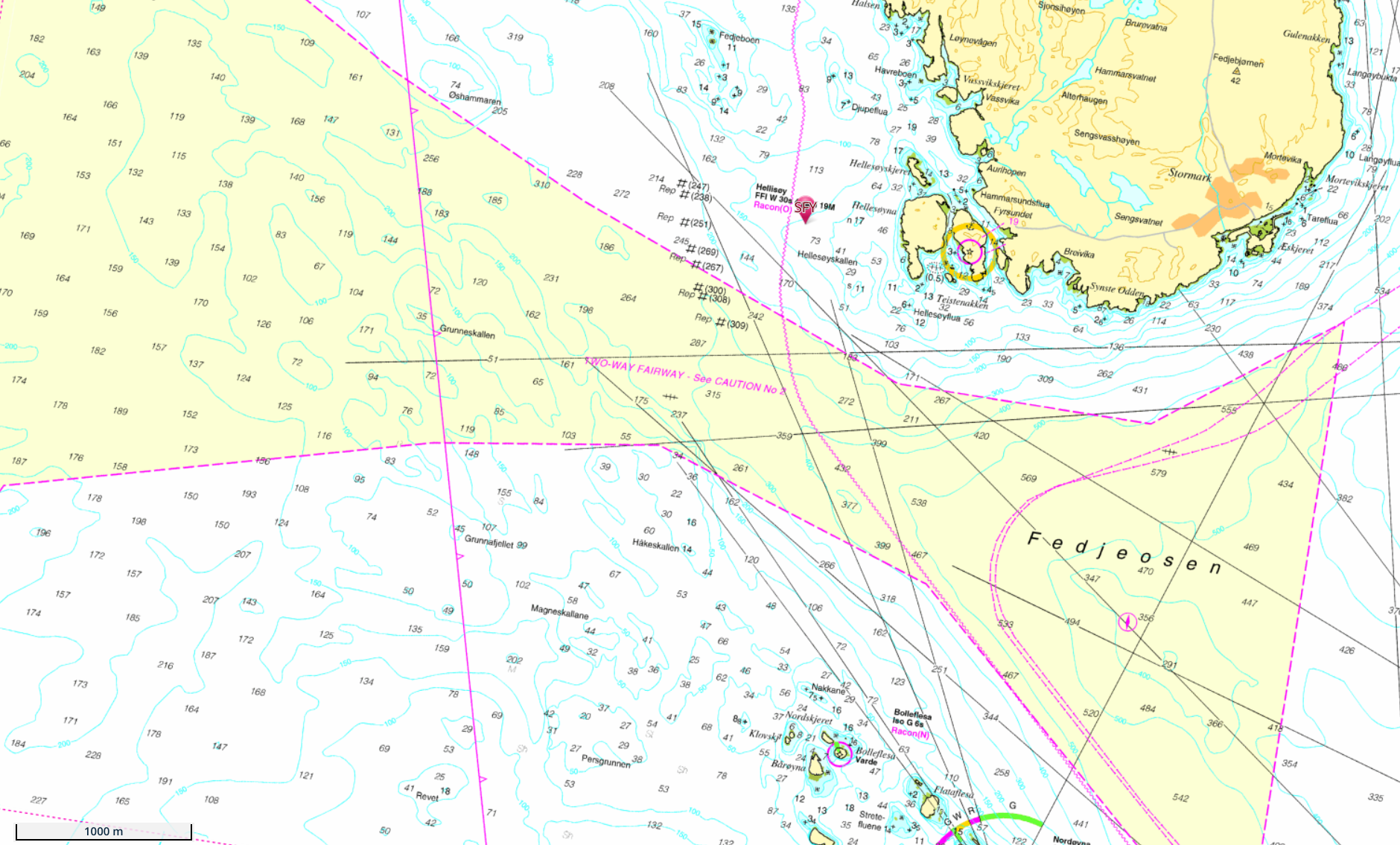}
  \caption{Location of mooring in Fedjeosen, map from Kartverket (sjøkart).}
  \label{fig:fedjeosen-map}
\end{figure*}

\begin{figure*}[ht]
  \centering
  \includegraphics[width=0.75\textwidth]{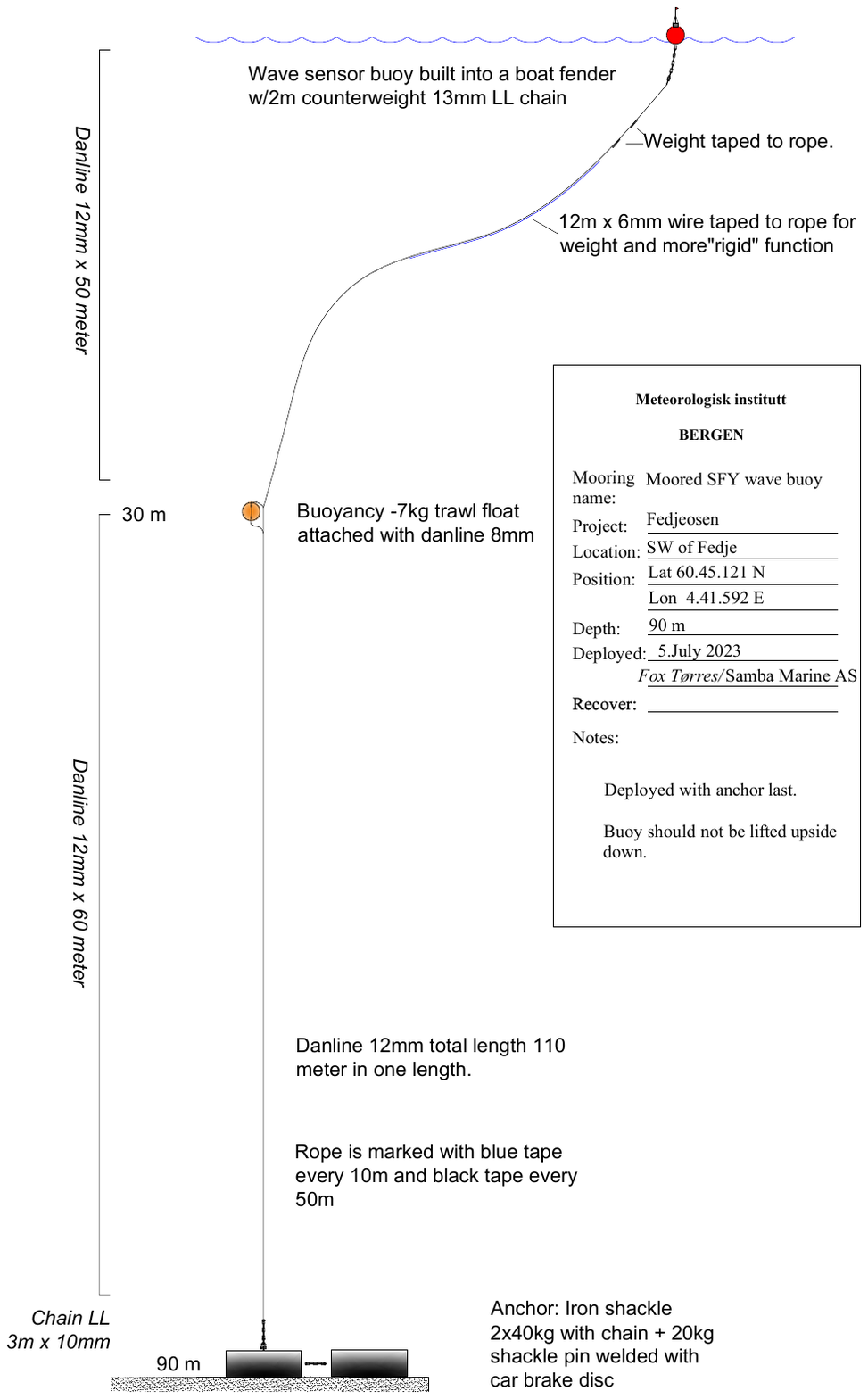}
  \caption{Schematic of mooring configuration in Fedjeosen.}
  \label{fig:fedjeosen-mooring-schematic}
\end{figure*}



\bibliographystyle{ametsocV6}
\bibliography{sfy}

\begin{thebibliography}{70}
\providecommand{\natexlab}[1]{#1}
\providecommand{\url}[1]{\texttt{#1}}
\renewcommand{\UrlFont}{\rmfamily}
\providecommand{\urlprefix}{URL }
\expandafter\ifx\csname urlstyle\endcsname\relax
  \providecommand{\doi}[1]{https://doi.org/\discretionary{}{}{}#1}\else
  \providecommand{\doi}{https://doi.org/\discretionary{}{}{}\begingroup
  \urlstyle{rm}\Url}\fi
\providecommand{\eprint}[2][]{\url{#2}}

\bibitem[{{Adafruit}(2023)}]{adafruit2023AdafruitAHRSLibrary}
{Adafruit}, 2023: Adafruit {{AHRS}} library. Adafruit Industries.

\bibitem[{Alari et~al.(2022)}]{alari2022LainePoissLightweightIceResistant}
Alari, V., and Coauthors, 2022: {{LainePoiss}}{\textregistered}---{{A
  Lightweight}} and {{Ice-Resistant Wave Buoy}}. \textit{J. Atmos. Oceanic
  Technol.}, \textbf{39~(5)}, 573--594, \doi{10.1175/JTECH-D-21-0091.1}.

\bibitem[{Ardhuin et~al.(2010)}]{ardhuin2010SemiempiricalDissipationSource}
Ardhuin, F., and Coauthors, 2010: Semiempirical {{Dissipation Source
  Functions}} for {{Ocean Waves}}. {{Part I}}: {{Definition}}, {{Calibration}},
  and {{Validation}}. \textit{Journal of Physical Oceanography},
  \textbf{40~(9)}, 1917--1941, \doi{10.1175/2010JPO4324.1}.

\bibitem[{Arraigada and Partl(2006)Arraigada, and
  Partl}]{arraigada2006CalculationDisplacementsMeasured}
Arraigada, M., and M.~Partl, 2006: Calculation of displacements of measured
  accelerations, analysis of two accelerometers and application in road
  engineering. \textit{6th {{Swiss Transport Research Conference}}}.

\bibitem[{Babanin(2009)}]{babanin2009BreakingOceanSurface}
Babanin, A., 2009: Breaking of ocean surface waves. \textit{Acta Physica
  Slovaca. Reviews and Tutorials}, \textbf{59~(4)},
  \doi{10.2478/v10155-010-0097-5}.

\bibitem[{Benetazzo(2006)}]{benetazzo2006MeasurementsShortWater}
Benetazzo, A., 2006: Measurements of short water waves using stereo matched
  image sequences. \textit{Coastal Engineering}, \textbf{53~(12)}, 1013--1032,
  \doi{10.1016/j.coastaleng.2006.06.012}.

\bibitem[{Bj{\"o}rkqvist et~al.(2016)Bj{\"o}rkqvist, Pettersson, Laakso, Kahma,
  Jokinen,, and Kosloff}]{bjorkqvist2016RemovingLowfrequencyArtefacts}
Bj{\"o}rkqvist, J.-V., H.~Pettersson, L.~Laakso, K.~K. Kahma, H.~Jokinen, and
  P.~Kosloff, 2016: Removing low-frequency artefacts from {{Datawell DWR-G4}}
  wave buoy measurements. \textit{Geoscientific Instrumentation, Methods and
  Data Systems}, \textbf{5~(1)}, 17--25, \doi{10.5194/gi-5-17-2016}.

\bibitem[{Boehm et~al.(2017)Boehm, Ismail, Sassoubre,, and
  Andruszkiewicz}]{boehm2017OceansPerilGrand}
Boehm, A.~B., N.~S. Ismail, L.~M. Sassoubre, and E.~A. Andruszkiewicz, 2017:
  Oceans in {{Peril}}: {{Grand Challenges}} in {{Applied Water Quality
  Research}} for the 21st {{Century}}. \textit{Environmental Engineering
  Science}, \textbf{34~(1)}, 3--15, \doi{10.1089/ees.2015.0252}.

\bibitem[{Bohlinger et~al.(2019)Bohlinger, Breivik, Economou,, and
  M{\"u}ller}]{bohlinger2019NovelApproachComputing}
Bohlinger, P., {\O}.~Breivik, T.~Economou, and M.~M{\"u}ller, 2019: A novel
  approach to computing super observations for probabilistic wave model
  validation. \textit{Ocean Modelling}, \textbf{139}, 101\,404,
  \doi{10.1016/j.ocemod.2019.101404}.

\bibitem[{Breivik and Allen(2008)Breivik, and
  Allen}]{breivik2008OperationalSearchRescue}
Breivik, {\O}., and A.~A. Allen, 2008: An {{Operational Search}} and {{Rescue
  Model}} for the {{Norwegian Sea}} and the {{North Sea}}. \textit{Journal of
  Marine Systems}, \textbf{69~(1-2)}, 99--113,
  \doi{10.1016/j.jmarsys.2007.02.010}, \eprint{1111.1102}.

\bibitem[{Brown et~al.(2018)Brown, Thomson, Ellenson, Rollano, {Ozkan-Haller},,
  and Haller}]{brown2018KinematicsStatisticsBreaking}
Brown, A., J.~Thomson, A.~Ellenson, F.~T. Rollano, H.~T. {Ozkan-Haller}, and
  M.~C. Haller, 2018: Kinematics and {{Statistics}} of {{Breaking Waves
  Observed Using SWIFT Buoys}}. \textit{IEEE J. Oceanic Eng.}, \textbf{44~(4)},
  1011--1023, \doi{10.1109/JOE.2018.2868335}.

\bibitem[{Brown and Paasch(2021)Brown, and
  Paasch}]{brown2021AccelerationsWaveMeasurement}
Brown, A.~C., and R.~K. Paasch, 2021: The {{Accelerations}} of a {{Wave
  Measurement Buoy Impacted}} by {{Breaking Waves}} in the {{Surf Zone}}.
  \textit{JMSE}, \textbf{9~(2)}, 214, \doi{10.3390/jmse9020214}.

\bibitem[{Chuang et~al.(2013)Chuang, Wu,, and
  Wang}]{chuang2013ContinuousWaveletTransform}
Chuang, L. Z.-H., L.-C. Wu, and J.-H. Wang, 2013: Continuous {{Wavelet
  Transform Analysis}} of {{Acceleration Signals Measured}} from a {{Wave
  Buoy}}. \textit{Sensors}, \textbf{13~(8)}, 10\,908--10\,930,
  \doi{10.3390/s130810908}.

\bibitem[{Collins et~al.(2014)Collins, Lund, Waseda,, and
  Graber}]{collins2014RecordingSeaSurface}
Collins, C.~O., B.~Lund, T.~Waseda, and H.~C. Graber, 2014: On recording sea
  surface elevation with accelerometer buoys: Lessons from {{ITOP}} (2010).
  \textit{Ocean Dynamics}, \textbf{64~(6)}, 895--904,
  \doi{10.1007/s10236-014-0732-7}.

\bibitem[{Collins et~al.(2024)}]{collins2024PerformanceMooredGPS}
Collins, C.~O., and Coauthors, 2024: Performance of moored {{GPS}} wave buoys.
  \textit{Coastal Engineering Journal}.

\bibitem[{{Datawell BV}(2023)}]{datawellbv2023DatawellWaveriderManual}
{Datawell BV}, 2023: Datawell {{Waverider Manual}} - {{DWR4}}. Tech. rep.

\bibitem[{Deike(2022)}]{deike2022MassTransferOcean}
Deike, L., 2022: Mass {{Transfer}} at the {{Ocean}}--{{Atmosphere Interface}}:
  {{The Role}} of {{Wave Breaking}}, {{Droplets}}, and {{Bubbles}}.
  \textit{Annual Review of Fluid Mechanics}, \textbf{54~(1)}, 191--224,
  \doi{10.1146/annurev-fluid-030121-014132}.

\bibitem[{Donelan et~al.(1996)Donelan, Drennan,, and
  Magnusson}]{donelan1996NonstationaryAnalysisDirectionala}
Donelan, M.~A., W.~M. Drennan, and A.~K. Magnusson, 1996: Nonstationary
  {{Analysis}} of the {{Directional Properties}} of {{Propagating Waves}}.
  \textit{Journal of Physical Oceanography}, \textbf{26~(9)}, 1901--1914,
  \doi{10.1175/1520-0485(1996)026<1901:NAOTDP>2.0.CO;2}.

\bibitem[{Donelan et~al.(1985)Donelan, Hamilton,, and
  Hui}]{donelan1985DirectionalSpectraWindgenerated}
Donelan, M.~A., J.~Hamilton, and W.~H. Hui, 1985: Directional spectra of
  wind-generated ocean waves. \textit{Phil. Trans. R. Soc. Lond. A},
  \textbf{315~(1534)}, 509--562, \doi{10.1098/rsta.1985.0054}.

\bibitem[{Dorgeville et~al.(2018)Dorgeville, Tholo,, and
  Tengberg}]{dorgeville2018DataValidationMOTUS}
Dorgeville, E., H.~Tholo, and D.~A. Tengberg, 2018: Data validation for the
  {{MOTUS Directional Wave Buoy}}. Tech. rep.

\bibitem[{Duracell(2019)}]{duracell2019DuracellProCellIntense}
Duracell, 2019: Duracell {{ProCell Intense}} ({{Datasheet}}). Tech. rep.

\bibitem[{{Edinburgh Designs}(2023)}]{edinburghdesigns2023PistonCoastalWave}
{Edinburgh Designs}, 2023: Piston {{Coastal Wave Generators}} {\textbar}
  {{Edinburgh Designs}}.
  http://www4.edesign.co.uk/product/piston-wave-generators/.

\bibitem[{{E.U. Copernicus Marine Service Information (CMEMS), Marine Data
  Store
  (MDS)}(2023)}]{e.u.copernicusmarineserviceinformationcmemsmarinedatastoremds2023GlobalOceanSignificant}
{E.U. Copernicus Marine Service Information (CMEMS), Marine Data Store (MDS)},
  2023: Global {{Ocean L}} 3 {{Significant Wave Height From Nrt Satellite
  Measurements}}. \doi{10.48670/moi-00179}.

\bibitem[{Feddersen et~al.(2023{\natexlab{a}})Feddersen, Amador, Pick, Vizuet,
  Quinn, Wolfinger, MacMahan,, and
  Fincham}]{feddersen2023WavedrifterLowcostIMUbased}
Feddersen, F., A.~Amador, K.~Pick, A.~Vizuet, K.~Quinn, E.~Wolfinger, J.~H.
  MacMahan, and A.~Fincham, 2023{\natexlab{a}}: The wavedrifter: A low-cost
  {{IMU-based Lagrangian}} drifter to observe steepening and overturning of
  surface gravity waves and the transition to turbulence. \textit{Coastal
  Engineering Journal}, 1--14, \doi{10.1080/21664250.2023.2238949}.

\bibitem[{Feddersen et~al.(2023{\natexlab{b}})Feddersen, Fincham, Brodie,
  Young, Spydell, Grimes, Pieszka,, and
  Hanson}]{feddersen2023CrossshoreWindinducedChanges}
Feddersen, F., A.~M. Fincham, K.~L. Brodie, A.~P. Young, M.~S. Spydell, D.~J.
  Grimes, M.~Pieszka, and K.~Hanson, 2023{\natexlab{b}}: Cross-shore
  wind-induced changes to field-scale overturning wave shape. \textit{Journal
  of Fluid Mechanics}, \textbf{958}, A4, \doi{10.1017/jfm.2023.40}.

\bibitem[{Fisher et~al.(2021)Fisher, Thomson,, and
  Schwendeman}]{fisher2021RapidDeterministicWave}
Fisher, A., J.~Thomson, and M.~Schwendeman, 2021: Rapid deterministic wave
  prediction using a sparse array of buoys. \textit{Ocean Engineering},
  \textbf{228}, 108\,871, \doi{10.1016/j.oceaneng.2021.108871}.

\bibitem[{Forristall(2000)}]{forristall2000WaveCrestDistributions}
Forristall, G.~Z., 2000: Wave {{Crest Distributions}}: {{Observations}} and
  {{Second-Order Theory}}. \textit{J. Phys. Oceanogr.}, \textbf{30~(8)},
  1931--1943, \doi{10.1175/1520-0485(2000)030<1931:WCDOAS>2.0.CO;2}.

\bibitem[{G{\"u}nther et~al.(1992)G{\"u}nther, Hasselmann,, and
  Janssen}]{gunther1992WAMModelCycle}
G{\"u}nther, H., S.~Hasselmann, and P.~A. E.~M. Janssen, 1992: The {{WAM Model
  Cycle}} 4. Tech. Rep.~4.

\bibitem[{Halsne(2024)}]{halsne2024WaveModulationStrong}
Halsne, T., 2024: Wave modulation by strong currents: {{A}} study in the
  {{Lofoten Maelstrom}} and surrounding areas. Doctoral thesis, The University
  of Bergen.

\bibitem[{Halsne et~al.(2023)Halsne, Benetazzo, Barbariol, Christensen,
  Carrasco,, and Breivik}]{halsne2023WaveModulationStrong}
Halsne, T., A.~Benetazzo, F.~Barbariol, K.~H. Christensen, A.~Carrasco, and
  {\O}.~Breivik, 2023: Wave modulation in a strong tidal current and its impact
  on extreme waves. \textit{Journal of Physical Oceanography},
  \doi{10.1175/JPO-D-23-0051.1}.

\bibitem[{Halsne et~al.(2022)Halsne, Bohlinger, Christensen, Carrasco,, and
  Breivik}]{halsne2022ResolvingRegionsKnown}
Halsne, T., P.~Bohlinger, K.~H. Christensen, A.~Carrasco, and {\O}.~Breivik,
  2022: Resolving regions known for intense wave--current interaction using
  spectral wave models: {{A}} case study in the energetic flow fields of
  {{Northern Norway}}. \textit{Ocean Modelling}, \textbf{176}, 102\,071,
  \doi{10.1016/j.ocemod.2022.102071}.

\bibitem[{Hasvold et~al.(2007)Hasvold, Forseth, Johannessen,, and
  Lian}]{hasvold2007SafetyAspectsLarge}
Hasvold, {\O}., S.~Forseth, T.~C. Johannessen, and T.~Lian, 2007: Safety
  aspects of large lithium batteries. Tech. rep., FFI.

\bibitem[{Holthuijsen(2010)}]{holthuijsen2010WavesOceanicCoastal}
Holthuijsen, L.~H., 2010: \textit{Waves in {{Oceanic}} and {{Coastal Waters}}}.
  Cambridge University Press.

\bibitem[{Komen et~al.(1996)Komen, Cavaleri, Donelan, Hasselmann, Hasselmann,,
  and Janssen}]{komen1996DynamicsModellingOcean}
Komen, G.~J., L.~Cavaleri, M.~Donelan, K.~Hasselmann, S.~Hasselmann, and P.~A.
  E.~M. Janssen, 1996: \textit{Dynamics and {{Modelling}} of {{Ocean Waves}}}.
  Cambridge University Press.

\bibitem[{{Longuet-Higgins}(1986)}]{longuet-higgins1986EulerianLagrangianAspects}
{Longuet-Higgins}, M.~S., 1986: Eulerian and {{Lagrangian}} aspects of surface
  waves. \textit{J. Fluid Mech.}, \textbf{173}, 683--707,
  \doi{10.1017/S0022112086001325}.

\bibitem[{Magnusson et~al.(1999)Magnusson, Donelan,, and
  Drennan}]{magnusson1999EstimatingExtremesEvolving}
Magnusson, A., M.~Donelan, and W.~Drennan, 1999: On estimating extremes in an
  evolving wave field. \textit{Coastal Engineering}, \textbf{36~(2)}, 147--163,
  \doi{10.1016/S0378-3839(99)00004-6}.

\bibitem[{Malila(2022)}]{malilamika2022InvestigationDynamicalStatistical}
Malila, M., 2022: An investigation into the dynamical and statistical
  properties of dominant ocean surface waves using close-range remote sensing.
  Ph.D. thesis, University of Bergen, Bergen.

\bibitem[{Malila et~al.(2022)Malila, Thomson, Breivik, Benetazzo, Scanlon,, and
  Ward}]{malila2022GroupinessIntermittencyOceanic}
Malila, M.~P., J.~Thomson, {\O}.~Breivik, A.~Benetazzo, B.~Scanlon, and
  B.~Ward, 2022: On the {{Groupiness}} and {{Intermittency}} of {{Oceanic
  Whitecaps}}. \textit{JGR Oceans}, \textbf{127~(1)}, e2021JC017\,938,
  \doi{10.1029/2021JC017938}.

\bibitem[{Martini et~al.(2021)Martini, Hoeke, Crossing, Groot, Branson,, and
  Pitman}]{martini2021MooringDesignOperational}
Martini, A., R.~Hoeke, R.~Crossing, B.~D. Groot, P.~Branson, and T.~Pitman,
  2021: Mooring {{Design}} and {{Operational Guidelines}} for {{Lightweight
  Wave Buoys}}. Tech. rep., CSIRO, Australia.

\bibitem[{Melville(1996)}]{melville1996RoleSurfaceWaveBreaking}
Melville, W.~K., 1996: The {{Role}} of {{Surface-Wave Breaking}} in {{Air-Sea
  Interaction}}. \textit{Annu. Rev. Fluid Mech.}

\bibitem[{M{\"u}ller
  et~al.(2017)}]{muller2017AROMEMetCoOpNordicConvectiveScale}
M{\"u}ller, M., and Coauthors, 2017: {{AROME-MetCoOp}}: {{A Nordic
  Convective-Scale Operational Weather Prediction Model}}. \textit{Weather and
  Forecasting}, \textbf{32~(2)}, 609--627, \doi{10.1175/WAF-D-16-0099.1}.

\bibitem[{Nose et~al.(2018)Nose, Webb, Waseda, Inoue,, and
  Sato}]{nose2018PredictabilityStormWavea}
Nose, T., A.~Webb, T.~Waseda, J.~Inoue, and K.~Sato, 2018: Predictability of
  storm wave heights in the ice-free {{Beaufort Sea}}. \textit{Ocean Dynamics},
  \textbf{68~(10)}, 1383--1402, \doi{10.1007/s10236-018-1194-0}.

\bibitem[{Nose et~al.(2023)}]{nose2023ComparisonOperationalWaveice}
Nose, T., and Coauthors, 2023: arXiv:2302.02820. A comparison of an operational
  wave-ice model product and drifting wave buoy observation in the central
  {{Arctic Ocean}}: Investigating the effect of sea ice forcing in thin ice
  cover. arXiv, \eprint{2302.02820}.

\bibitem[{{NXP Semiconductors}(2016)}]{nxpsemiconductors2016NXPSensorFusion}
{NXP Semiconductors}, 2016: {{NXP Sensor Fusion}} - {{NXP Sensor Fusion}} for
  {{Kinetis MCUs}}.

\bibitem[{Rabault et~al.(2020)Rabault, Sutherland, Gundersen, Jensen,
  Marchenko,, and Breivik}]{rabault2020OpenSourceVersatile}
Rabault, J., G.~Sutherland, O.~Gundersen, A.~Jensen, A.~Marchenko, and
  {\O}.~Breivik, 2020: An open source, versatile, affordable waves in ice
  instrument for scientific measurements in the {{Polar Regions}}. \textit{Cold
  Regions Science and Technology}, \textbf{170}, 102\,955,
  \doi{10.1016/j.coldregions.2019.102955}.

\bibitem[{Rabault
  et~al.(2022)}]{rabault2022OpenMetBuoyv2021EasytoBuildAffordable}
Rabault, J., and Coauthors, 2022: {{OpenMetBuoy-v2021}}: {{An Easy-to-Build}},
  {{Affordable}}, {{Customizable}}, {{Open-Source Instrument}} for
  {{Oceanographic Measurements}} of {{Drift}} and {{Waves}} in {{Sea Ice}} and
  the {{Open Ocean}}. \textit{Geosciences}, \textbf{12~(3)}, 110,
  \doi{10.3390/geosciences12030110}.

\bibitem[{Rabault et~al.(2023)}]{rabault2023DatasetDirectObservations}
Rabault, J., and Coauthors, 2023: A dataset of direct observations of sea ice
  drift and waves in ice. \textit{Sci. Data}.

\bibitem[{Raghukumar et~al.(2019)Raghukumar, Chang, Spada, Jones, Janssen,, and
  Gans}]{raghukumar2019PerformanceCharacteristicsSpotter}
Raghukumar, K., G.~Chang, F.~Spada, C.~Jones, T.~Janssen, and A.~Gans, 2019:
  Performance {{Characteristics}} of ``{{Spotter}},'' a {{Newly Developed
  Real-Time Wave Measurement Buoy}}. \textit{Journal of Atmospheric and Oceanic
  Technology}, \textbf{36~(6)}, 1127--1141, \doi{10.1175/JTECH-D-18-0151.1}.

\bibitem[{Rainville et~al.(2023)Rainville, Thomson, Moulton,, and
  Derakhti}]{rainville2023MeasurementsNearshoreOceansurface}
Rainville, E., J.~Thomson, M.~Moulton, and M.~Derakhti, 2023: Measurements of
  nearshore ocean-surface kinematics through coherent arrays of free-drifting
  buoys. \textit{Earth System Science Data}, \textbf{15~(11)}, 5135--5151,
  \doi{10.5194/essd-15-5135-2023}.

\bibitem[{Rainville(2022)}]{rainville2022MeasurementsNearshoreWaves}
Rainville, E.~J., 2022: Measurements of {{Nearshore Waves}} through {{Coherent
  Arrays}} of {{Small-Scale}}, {{Free-Drifting Wave Buoys}}. {{MSc}},
  University of Washington, Seattle.

\bibitem[{Saetra et~al.(2021)Saetra, Halsne, Carrasco, Breivik, Pedersen,, and
  Christensen}]{saetra2021IntenseInteractionsOcean}
Saetra, {\O}., T.~Halsne, A.~Carrasco, {\O}.~Breivik, T.~Pedersen, and K.~H.
  Christensen, 2021: Intense interactions between ocean waves and currents
  observed in the {{Lofoten Maelstrom}}. \textit{Journal of Physical
  Oceanography}, \doi{10.1175/JPO-D-20-0290.1}.

\bibitem[{Saetre et~al.(2023)}]{saetre2023DirectionalWaveMeasurements}
Saetre, C., and Coauthors, 2023: Directional wave measurements from
  navigational buoys. \textit{Ocean Engineering}, \textbf{268}, 113\,161,
  \doi{10.1016/j.oceaneng.2022.113161}.

\bibitem[{Schwendeman and Thomson(2014)Schwendeman, and
  Thomson}]{schwendeman2014WaveBreakingDissipation}
Schwendeman, M., and J.~Thomson, 2014: Wave {{Breaking Dissipation}} in a
  {{Young Wind Sea}} in: {{Journal}} of {{Physical Oceanography}}.
  \textit{Journal of Physical Oceanography}, \textbf{44~(1)}.

\bibitem[{Seldal(2023)}]{seldal2023SFYFreedriftingWave}
Seldal, T.~I., 2023: {{SFY}}: {{A}} free-drifting wave buoy for measuring surf
  waves. M.S. thesis, The University of Bergen.

\bibitem[{Sinclair(2014)}]{sinclair2014FlowRiderLagrangianFloat}
Sinclair, A., 2014: {{FlowRider}}: {{A Lagrangian Float}} to {{Measure}} 3-{{D
  Dynamics}} of {{Plunging Breakers}} in the {{Surf Zone}}. \textit{Journal of
  Coastal Research}, \textbf{293}, 205--209,
  \doi{10.2112/JCOASTRES-D-13-00014.1}.

\bibitem[{Singsaas et~al.(2020)Singsaas, Leirvik, Daling, Gu{\'e}nette,, and
  S{\o}rheim}]{singsaas2020FateBehaviourWeathered}
Singsaas, I., F.~Leirvik, P.~S. Daling, C.~Gu{\'e}nette, and K.~R. S{\o}rheim,
  2020: Fate and behaviour of weathered oil drifting into sea ice, using a
  novel wave and current flume. \textit{Marine Pollution Bulletin},
  \textbf{159}, 111\,485, \doi{10.1016/j.marpolbul.2020.111485}.

\bibitem[{{STMicroelectronics}(2019)}]{stmicroelectronics2019LSM6DSOXDatasheet}
{STMicroelectronics}, 2019: {{LSM6DSOX}} ({{Datasheet}}). Tech. rep.

\bibitem[{STMicroelectronics(2020)}]{stmicroelectronics2020ISM330DHCXDatasheet}
STMicroelectronics, 2020: {{ISM330DHCX}} ({{Datasheet}}). Tech. rep.

\bibitem[{{STMicroelectronics}(2021)}]{stmicroelectronics2021AN5398ISM330DHCXAlwayson}
{STMicroelectronics}, 2021: {{AN5398}}: {{ISM330DHCX}}: Always-on {{3D}}
  accelerometer and {{3D}} gyroscope with digital output for industrial
  applications.

\bibitem[{{The Wamdi Group}(1988)}]{thewamdigroup1988WAMModelThird}
{The Wamdi Group}, 1988: The {{WAM Model}}---{{A Third Generation Ocean Wave
  Prediction Model}}. \textit{Journal of Physical Oceanography},
  \textbf{18~(12)}, 1775--1810,
  \doi{10.1175/1520-0485(1988)018<1775:TWMTGO>2.0.CO;2}.

\bibitem[{Thomson(2012)}]{thomson2012WaveBreakingDissipation}
Thomson, J., 2012: Wave {{Breaking Dissipation Observed}} with ``{{SWIFT}}''
  {{Drifters}}. \textit{Journal of Atmospheric and Oceanic Technology},
  \textbf{29~(12)}, 1866--1882, \doi{10.1175/JTECH-D-12-00018.1}.

\bibitem[{Thomson et~al.(2018)Thomson, Girton, Jha,, and
  Trapani}]{thomson2018MeasurementsDirectionalWave}
Thomson, J., J.~B. Girton, R.~Jha, and A.~Trapani, 2018: Measurements of
  {{Directional Wave Spectra}} and {{Wind Stress}} from a {{Wave Glider
  Autonomous Surface Vehicle}}. \textit{Journal of Atmospheric and Oceanic
  Technology}, \textbf{35~(2)}, 347--363, \doi{10.1175/JTECH-D-17-0091.1}.

\bibitem[{Thomson et~al.(2015)}]{thomson2015BiofoulingEffectsResponse}
Thomson, J., and Coauthors, 2015: Biofouling {{Effects}} on the {{Response}} of
  a {{Wave Measurement Buoy}} in {{Deep Water}}. \textit{Journal of Atmospheric
  and Oceanic Technology}, \textbf{32~(6)}, 1281--1286,
  \doi{10.1175/JTECH-D-15-0029.1}.

\bibitem[{Thomson et~al.(2019)}]{thomson2019NewVersionSWIFT}
Thomson, J., and Coauthors, 2019: A new version of the {{SWIFT}} platform for
  waves, currents, and turbulence in the ocean surface layer. \textit{2019
  {{IEEE}}/{{OES Twelfth Current}}, {{Waves}} and {{Turbulence Measurement}}
  ({{CWTM}})}, IEEE, San Diego, CA, USA, 1--7,
  \doi{10.1109/CWTM43797.2019.8955299}.

\bibitem[{Thomson et~al.(2023)}]{thomson2023DevelopmentTestingMicroSWIFTa}
Thomson, J., and Coauthors, 2023: Development and testing of {{microSWIFT}}
  expendable wave buoys. \textit{Coastal Engineering Journal}, 1--13,
  \doi{10.1080/21664250.2023.2283325}.

\bibitem[{Tucker(1958)}]{tucker1958AccuracyWaveMeasurementsb}
Tucker, M., 1958: The accuracy of wave measurements made with vertical
  accelerometers. \textit{Deep Sea Research (1953)}, \textbf{5~(2-4)},
  185--192, \doi{10.1016/0146-6313(58)90009-1}.

\bibitem[{Van~Essen et~al.(2018)Van~Essen, Ewans,, and
  McConochie}]{vanessen2018WaveBuoyPerformance}
Van~Essen, S., K.~Ewans, and J.~McConochie, 2018: Wave {{Buoy Performance}} in
  {{Short}} and {{Long Waves}}, {{Evaluated Using Tests}} on a {{Hexapod}}.
  \textit{Volume {{7B}}: {{Ocean Engineering}}}, American Society of Mechanical
  Engineers, Madrid, Spain, V07BT06A001, \doi{10.1115/OMAE2018-77092}.

\bibitem[{Veras~Guimar{\~a}es
  et~al.(2018)}]{verasguimaraes2018SurfaceKinematicsBuoy}
Veras~Guimar{\~a}es, P., and Coauthors, 2018: A surface kinematics buoy
  ({{SKIB}}) for wave--current interaction studies. \textit{Ocean Sci.},
  \textbf{14~(6)}, 1449--1460, \doi{10.5194/os-14-1449-2018}.

\bibitem[{Waseda et~al.(2018)Waseda, Webb, Sato, Inoue, Kohout, Penrose,, and
  Penrose}]{waseda2018CorrelatedIncreaseHigh}
Waseda, T., A.~Webb, K.~Sato, J.~Inoue, A.~Kohout, B.~Penrose, and S.~Penrose,
  2018: Correlated {{Increase}} of {{High Ocean Waves}} and {{Winds}} in the
  {{Ice-Free Waters}} of the {{Arctic Ocean}}. \textit{Sci Rep},
  \textbf{8~(1)}, 4489, \doi{10.1038/s41598-018-22500-9}.

\bibitem[{Yurovsky and Dulov(2020)Yurovsky, and
  Dulov}]{yurovsky2020MEMSbasedWaveBuoya}
Yurovsky, Y.~Y., and V.~A. Dulov, 2020: {{MEMS-based}} wave buoy: {{Towards}}
  short wind-wave sensing. \textit{Ocean Engineering}, \textbf{217}, 108\,043,
  \doi{10.1016/j.oceaneng.2020.108043}.

\end{thebibliography}

\end{document}